\documentclass[11pt,english,floatfix,superscriptaddress,aps,prd,onecolumn,preprint,showpacs,showkeys]{revtex4}
\usepackage{amsmath}
\usepackage{amssymb}
\usepackage{amsbsy}
\usepackage{amsfonts}
\usepackage{amsopn}
\usepackage{amstext}
\usepackage{graphicx}
\usepackage{amssymb}
\usepackage{amsfonts}
\usepackage{amsmath}
\usepackage{graphicx}
\usepackage[english]{babel}
\usepackage{color}
\usepackage{esint}
\usepackage[dvips]{epsfig}
\usepackage[dvips]{graphicx}
\usepackage{float}
\usepackage{units}
\usepackage{textcomp}

\usepackage{hyperref}
\hypersetup{
    colorlinks,
    citecolor=blue,
    filecolor=red,
    linkcolor=blue,
    urlcolor=blue
}

\usepackage{hyperref}
\usepackage{slashed}

\usepackage{graphicx} 
\usepackage{subfigure} 

\DeclareMathOperator{\e}{e}
\DeclareMathOperator{\sech}{sech}

\DeclareMathOperator{\sn}{sn}
\DeclareMathOperator{\cn}{cn}
\DeclareMathOperator{\dn}{dn}

\begin{document}

\title{Corrections to Newton's law of gravitation in the context of codimension-$1$ warped thick braneworlds}

\author{D. F. S. Veras}
\email{franklin@fisica.ufc.br}

\author{C. A. S. Almeida}
\email{carlos@fisica.ufc.br}

\affiliation{Departamento de F\'{i}sica, Universidade Federal do Cear\'{a} (UFC), Campus do Pici, Caixa Postal 6030, 60455-760, Fortaleza, Cear\'{a}, Brazil}


\begin{abstract}
In this work, we compute the corrections in the Newton's law of gravitation due to Kaluza-Klein gravitons in codimension-$1$ warped thick braneworld scenarios. We focus in some models recently proposed in the literature, the so-called asymmetric hybrid brane and compact brane. Such models are deformations of the $\phi^4$ and sine-Gordon topological defects, respectively. Therefore we consider the branes engendered by such defects and we also compute the corrections in their cases. We use suitable numerical techniques to attain the mass spectrum and its corresponding eigenfunctions which are the essential quantities for computing the correction to the Newtonian potential. Moreover, we discuss that the existence of massive modes is necessary for building a braneworld model with a phenomenology involved. We find that the odd eigenfunctions have non-trivial contributions and the first eigenstate of the Kaluza-Klein tower has the highest contribution. The calculation of slight deviations in the gravitational potential may be used as a selection tool for braneworld scenarios matching with future experimental measurements in high energy collisions.

\end{abstract}


\maketitle


\section{Introduction}
\label{Sec_Introduction}

The idea that our observable universe may be thought as a $(3+1)$-hypersurface (called \textit{brane} or \textit{membrane}) embedded in a higher-dimensional bulk space-time has received a lot of attention since it provides good explanations to several puzzling phenomena  such as the hierarchy between the electroweak and the Planck scales \cite{RS1, ADJ}, the dark matter origin \cite{ADDK}, the cosmological constant problem \cite{CosmologicalProblem} and the cosmic acceleration \cite{CosmicAccel}. The braneworld concept has emerged from string theory \cite{Horava-Witten} and unification models \cite{ADD, AADD} and therefore, tests of the existence of branes, or just extra dimensions, are a fundamental challenge. Evidences of the braneworld hypothesis are expected to be found through high energy collisions (for instance in the Large Hadron Collider) wherein new degrees of freedom should arise such as Kaluza-Klein (KK) excitation of the particles in the extra dimension \cite{Hooper-Profumo, Branons}. 


Besides the search for extra dimension evidences in high energy particle colliders, it could be also observed in low-energy experiments using neutrons if a second brane exists close enough to us \cite{Sarrazin-Petit-PLB-2012}. Neutrons are more suitable than electrons, protons or atoms for such purpose \cite{Sarrazin-Petit-IJMPA}. Indeed, Dubovsky \textit{et al}  have suggested that the particles of the standard model should be able to leak into the bulk through a tunneling effect \cite{ParticleEscape1, ParticleEscape2}. Furthermore, Sarrazin and Petit have shown that for a bulk containing at least two parallel $3$-branes hidden to each other, matter swapping between these two worlds should occur triggered by the use of suitable ambient magnetic vector potentials \cite{Sarrazin-Petit-PRD-2010, Sarrazin-Petit-IJMPA, Sarrazin-Petit-PLB-2005, Sarrazin-Petit-PRD-2011}. More important, this new effect could be detected and controlled with present day technology as a possible experimental confirmation of the braneworld hypothesis. Recently, it was proposed an experiment to investigate matter swapping between branes by looking at the appearance of neutrons in a concept similar to \textit{light-shining-through-a-wall} \cite{Sarrazin-Petit-PRD-2015}. Ultracold neutrons stored in a vessel would therefore have a non-zero probability to escape from our brane toward the hidden brane at each wall collision. Such a process would be perceived as a neutron disappearance from the point of view of an observer in the visible brane \cite{Sarrazin-Petit-PLB-2012}. At last, it is worthwhile to mention the quite recently work by Yu and collaborators \cite{yu-liu}, which proposed that gravitational wave observations of compact binaries and their possible electromagnetic counterparts may be used to probe the nature of the extra dimension.

In this work, we will follow the original proposal of the phenomenological consequences of the braneworld hypothesis: slight deviations in the Newtonian potential between two particles \cite{RS2}. In order to have an effective four-dimensional theory of gravity, it is required that the models admit a normalizable zero-energy ground state which mediates the four-dimensional gravity \cite{Csaki-UniversalAspects}. We, in turn, discuss the importance of a model supporting massive states to have a phenomenology involved. In general, such states can not be found analytically. We, therefore, use suitable numerical techniques to attain the mass spectrum and its corresponding eigenvalues to compute the corrections in the Newton's law of gravitation for two braneworld models recently proposed in the literature, the so-called asymmetric hybrid brane \cite{AsymmetricHybridBrane} and the compact brane \cite{CompactBrane}. As desired, the spectra are real and monotonically increasing. Moreover, as the parameters of the models (generally related to the brane tension) increase, the complete spectra decrease. We show that the compact brane does not support massive states and, consequently, does not have contributions to corrections in the Newton's law. Moreover, the asymmetric hybrid brane has a negligible correction. Since these sophisticated braneworld scenarios use deformations from the $\phi^4$ and sine-Gordon topological defects, respectively, we study the phenomenology in branes engendered by such defects. In this new scenario we find that the sine-Gordon brane \cite{sine-GordonBrane} has higher contribution than the $\phi^4$ brane. In a previous work \cite{DiegoHybrid}, we studied the gravity localization in a recent braneworld model, called symmetric hybrid brane \cite{Bazeia-FromKinksToCompactons} and computed its corrections to the Newton's law. This procedure is very useful as a selection tool for brane models to match with future experimental measurements. As a matter of fact, it is important to mention that researches on corrections in Newton's law is a very attractive area in High Energy Physics as a theoretical foundation. For example, corrections to Newtonian potential in Palatini $f(R)$ braneworld scenario was performed using an approximate analytic approach \cite{Palatini}. Moreover, a quantum corrected Newton's law was derived in a loop quantum black holes context \cite{CarlosAlex} .

This paper is organized as follows: In Sec. \ref{Sec_ThickBrane}, we briefly review the general procedure for building a codimension-$1$ warped thick braneworld scenario and its metric fluctuations (also called gravity localization). We also discuss about the massive spectrum and the corrections to the Newton's law of gravitation. In the Sec. \ref{Model 1} and \ref{Model 2}, we present the main features of the asymmetric hybrid brane and compact brane models, respectively. We, therefore, reinforce their main issues analyzing the scalar curvature and the analogue quantum potential in details. In the Sec. \ref{Sec_Results}, we present our numerical results. In order to ensure the stability of our routines, we firstly apply the numerical methods in the Randall-Sundrum model and show that the relative error between the numerical results and the analytical solutions is less than $0.5\%$ in the first twenty eigenvalues. In the Sec. \ref{Sec-Discussion}, we discuss the results and the conclusions are given in the Sec. \ref{Sec_Conclusion}.



\section{Thick brane models and metric fluctuations}
\label{Sec_ThickBrane}
We will now present a brief review of the usual procedure for building warped codimension-$1$ thick braneworlds. Consider an auxiliary scalar field $\phi$ coupled to gravity in a five-dimensional warped space-time with an extra coordinate $y$ of infinite extent. The geometry is described by the metric
\begin{equation}
\label{metric}
ds^2 = \e^{2A(y)}\eta_{\mu\nu}dx^{\mu}dx^{\nu} - dy^2, 
\end{equation}
where $A(y)$ is the warp function and $\eta_{\mu\nu}$ is the Minkowski metric. The general Einstein-Hilbert action with matter in $D$ dimensions is given by \cite{Oda}
\begin{equation}
S_D = - \frac{1}{2{\kappa_D}^2}\int d^Dx \, \sqrt{-g} \, R + \int d^Dx \, \sqrt{g} \, \mathcal{L},
\end{equation}
where $\kappa_D$ denotes the $D$-dimensional gravitational constant related to the $D$-dimensional Newton constant $G_N$ and the $D$-dimensional Planck mass scale as 
\begin{equation}
{\kappa_D}^2 = 8\pi G_N = \frac{8\pi}{{M_{*}}^{D-2}}.
\label{NewtonianConstant}
\end{equation}
Moreover, $R$ is the Ricci scalar and $\mathcal{L}(\phi,\partial_M\phi)$ is the Lagrangian density which describes the scalar field
\begin{equation}
\mathcal{L}(\phi,\partial_M\phi) = \partial_M\phi\partial^M\phi - V(\phi),
\end{equation}
where $M = 0, 1,\cdots,D-1$. The function $V(\phi)$ is the potential of the model. The usual $\phi^4$ and sine-Gordon potentials, which have kink-like solutions, conduct to branes modeled by a domain wall defect \cite{Gremm,sine-GordonBrane}. Moreover, after the deformation procedure developed in the Ref. \cite{DeformedDefects}, different braneworld scenarios could be built using several distinct topological defects, engendering branes with richer internal structures \cite{Branalizacao, BlochBrane, Multikinks, Method-Obtaining}.

Henceforth, we will work in $D = 5$ in the notation $4\pi G_5 = 1$ of the Refs. \cite{AsymmetricHybridBrane, CompactBrane}. Supposing that the scalar field only depends on the extra dimension, the equations of motion, together with the Einstein equations, read  \cite{Gremm}
\begin{equation}
\label{Eq-mov1}
\phi^{\prime\prime} + 4\phi^{\prime}A^{\prime} = \frac{dV}{d\phi},
\end{equation}
\begin{equation}
\label{Eq-mov2}
A^{\prime\prime} = -\frac{2}{3} \left( \phi^{\prime} \right)^2,
\end{equation}
\begin{equation}
\label{Eq-mov3}
\left( A^{\prime} \right)^2 = \frac{1}{6} \phi^{\prime} - \frac{1}{3}V(\phi),
\end{equation}
where the primes denote derivatives with respect to $y$. The general treatment of the metric fluctuations is rather difficult, since one has to solve an intricate system of coupled non-linear differential equations \cite{Critical}. On the other hand, the superpotential approach \cite{Branalizacao} is an useful mechanism to reduce the second order equations of motion to first order ones. To this aim, the superpotential $W(\phi)$ is defined as \cite{Branalizacao}
\begin{equation}
V(\phi) = \frac{1}{2} \left( \frac{dW}{d\phi}\right)^2.
\label{V(phi)=dWdphi}
\end{equation}
Hence, the first order differential equations
\begin{equation}
\label{FirstOrder_EqMotion}
\phi^{\prime} = \frac{dW}{d\phi}, \hspace{0.5cm} \text{and} \hspace{0.5cm}  A^{\prime} = -\frac{2}{3}W(\phi) 
\end{equation}
solve the equations of motion \eqref{Eq-mov1}, \eqref{Eq-mov2} and \eqref{Eq-mov3}.  Therefore, the potential in the curved space-time reads
\begin{equation}
\label{Potential-Brane}
\mathcal{V}(\phi) = \frac{1}{2} \left( \frac{dW}{d\phi} \right)^2 - \, \frac{4}{3} W^2(\phi).
\end{equation}
The stability of a braneworld scenario is studied by means of the gravity localization. The usual mechanism to study the metric fluctuations is performing the following perturbation $h_{\mu\nu}(\mathbf{x},y)$ \cite{Gremm}
\begin{equation}
\label{perturbation}
ds^2 = \e^{2A(y)}(\eta_{\mu\nu} + h_{\mu\nu})dx^{\mu\nu}dx_{\mu\nu} + dy^2.
\end{equation}
Imposing the transverse-traceless gauge, i.e., $\partial_{\mu}h^{\mu\nu} = h_{\mu}^{\mu} = 0$ and $h_{5N} = 0$, the graviton equation of motion is \cite{Gremm}
\begin{equation}
\label{Graviton-Sturm}
h_{\mu\nu}^{\prime\prime} + \frac{ 2\sigma^{\prime}}{\sigma}h_{\mu\nu}^{\prime} = \sigma^{-1} \Box h_{\mu\nu},
\end{equation}
where $\Box$ is the $(3+1)-$dimensional d'Alembertian and $\sigma(y) = \e^{2A(y)}$. Furthermore, assuming the Kaluza-Klein (KK) decomposition $h_{\mu\nu}(\mathbf{x},y) = \sum_n h_{\mu\nu}^{(0)}(\mathbf{x})\varphi_n(y)$, where $\eta_{\mu\nu}\partial^{\mu}\partial^{\nu}h_{\mu\nu}^{(0)} = - {m_n}^2 h_{\mu\nu}^{(0)}$, with $m_n$ being the four-dimensional KK masses of the fluctuation, the gravitational KK modes in the extra dimension are described by the following Sturm-Liouville equation
\begin{equation}
\label{Graviton-Sturm-ExtraCoord}
\varphi_n^{\prime\prime}(y) + \frac{ 2\sigma^{\prime}}{\sigma}\varphi_n^{\prime}(y) = - {m_n}^2\sigma^{-1} \varphi_n(y).
\end{equation}

It is convenient to deal with a conformal metric performing the change of coordinates $dz = \sigma^{-1/2} dy$ \cite{Metastable}. Further, defining $\varphi_n(y) =  \sigma^{-3/4}\psi_n(z)$,  the Sturm-Liouville equation \eqref{Graviton-Sturm-ExtraCoord} reduces to a Schr\"{o}dinger-like form \cite{Metastable}
\begin{equation}
\label{Graviton-Schr}
-\ddot{\psi_n}(z) + U(z)\psi_n(z) = {m_n}^2 \psi_n(z), 
\end{equation}
where
\begin{equation}
\label{Potencial-Schr}
U(z)= \frac{3}{4}\left[ \frac{\ddot{\sigma}}{\sigma} - \frac{1}{4} \left(\frac{\dot{\sigma}}{\sigma}\right)^2\right],
\end{equation}
is the \textit{analogue quantum potential} and the over-dots represents derivatives with respect to $z$. A necessary condition to keep the stability of the gravitational sector is the absence of tachyonic states. This is ensured by the Hamiltonian can be written as \cite{Csaki-UniversalAspects}
\begin{equation}
H = \left\{ - \frac{d}{dz} + \frac{3}{4}\frac{\dot{\sigma}}{\sigma} \right\}\left\{ \frac{d}{dz} + \frac{3}{4}\frac{\dot{\sigma}}{\sigma} \right\},
\end{equation}
which has a supersymmetric quantum mechanics form 
\begin{equation}
Q^{\dagger}Q\psi_n(z) = m^2\psi_n(z), \hspace{1cm} \text{with} \hspace{1cm} Q \equiv \frac{d}{dz} + \frac{3}{4}\frac{\dot{\sigma}}{\sigma}.
\end{equation}
Furthermore, the Eq. \eqref{Graviton-Schr} has a zero-mode (massless graviton) solution for a generic form of the warp factor \cite{Metastable}
\begin{equation}
\psi_0(z) = \sigma^{3/2}(z),
\end{equation} 
which is normalizable. This state reproduces the four-dimensional gravity \cite{Metastable}.

An important geometric entity is the scalar curvature $R(y)$ which contains essential information. It is given in terms of the warp factor as
\begin{equation}
R(y) = - \frac{1}{4}\left[ \frac{\sigma^{\prime\prime}}{\sigma} + 19 \left(\frac{\sigma^{\prime}}{\sigma}\right)^2\right].
\label{ScalarCurvature}
\end{equation}
The $AdS_5$ limit of the bulk is characterized by the asymptotic behavior of the scalar curvature tending to a negative constant. Moreover, the presence of regions with positive scalar Ricci can, in principle, be connected with the capability to trap massive modes near the brane as resonant states \cite{Wilami-Resonance-Deformed}. 



\subsection{Graviton Spectrum}

The masses of the Kaluza-Klein (KK) tower are the most important quantities to study the phenomenology of a braneword scenario, since it is used directly to compute the corrections in the Newtonian potential. In general, the masses of the graviton excitations can not be obtained analytically. Nonetheless, in the Randall-Sundrum model (RS1), the  spectrum has a closed form. In this work, we are interested in the KK masses for the thick branes developed in the Refs. \cite{AsymmetricHybridBrane, CompactBrane}. But firstly, we will present the analytical result for the thin brane to test the efficiency of our routines.

The general solution of the Eq. \eqref{Graviton-Schr} in the RS scenario, whose warp factor is $\sigma = \e^{-2k|y|}$, is a linear combination of Bessel functions \cite{RS2}: 
\begin{equation}
\psi_n(z) = \sqrt{|z| + 1/k}\Bigg[a_n J_2\Big(m_n\left(|z|+1/k\right)\Big) + b_n Y_2\Big(m_n\left(|z|+1/k\right)\Big) \Bigg].
\label{RS-MassiveModes}
\end{equation}
The presence of two branes, which corresponds to a cut-off in the extra dimension, induces the quantization of the masses of the Kaluza-Klein states. The spectrum is obtained imposing the boundary conditions \cite{RS2}:
\begin{equation}
\psi_n^{\prime}(L_z) = - \frac{3k^2}{2(kL_z + 1)} \psi_n(L_z),
\label{RS-BoundaryConditions}
\end{equation}
where $L_z$ is the location of the TeV brane in the $z-$coordinate ($L_z = 0$ corresponds to the Planck brane). The transformation of coordinates $z(y)$ has the analytical solution for the RS warp factor: $|z| = (\e^{k|y|} - 1)/k$. Then, returning to the coordinate $y$, which effectively represents the distance along the extra dimension, the cut-off turns to $L_z = \e^{kL}/k$. Hence, imposing the boundary conditions \eqref{RS-BoundaryConditions} in the general solution \eqref{RS-MassiveModes}, we have
\begin{equation}
\begin{split}
a_n J_1 \left(\frac{m_n}{k}\right) + b_n Y_1 \left(\frac{m_n}{k}\right) & = 0,\\
a_n J_1 \left(\frac{m_n}{k}\e^{kL}\right) + b_n Y_1 \left(\frac{m_n}{k}\e^{kL}\right) & = 0,
\end{split}
\end{equation}
which has solutions if its determinant vanishes, i.e.:
\begin{equation}
\frac{J_1 \left(\frac{m_n}{k}\right)}{J_1 \left(\frac{m_n}{k}\e^{kL}\right)} = \frac{Y_1 \left(\frac{m_n}{k}\right)}{Y_1 \left(\frac{m_n}{k}\e^{kL}\right)}.
\end{equation}
In the approximation of \textit{small masses}, where $m_n/k \ll 1$, we have
\begin{equation}
J_1 \left(\frac{m_n}{k}\e^{kL}\right) = 0.
\end{equation}
Therefore, the masses of the Kaluza-Klein tower are given by
\begin{equation}
m_n \approx k\e^{-kL}j_n,
\label{RS-Spectrum}
\end{equation}
where $j_n$ are the zeroes of the Bessel function $J_1(j_n) = 0$ and are all tabled. The spectrum is discrete and monotonically increasing, but not equally spaced. Moreover, in the RS1 model, the spectrum reduces to $m_n = kj_n \e^{-kr_c\pi}$ \cite{Phenomenology-RS}, where $r_c$ is the compactification radius of the extra dimension. Furthermore, there is an exponential suppressed mass gap between the massless mode and the first massive one. It is important to note that the cut-off has influence in the spectrum. Then, future observations of deviations in Newtonian potential may be used to adjust the parameters $L$ and $k$.


\subsection{Corrections in Newton's law}

In order to have localized four-dimensional gravity, it is also required that the other solutions of the Schr\"odinger-like equation, the Kauza-Klein modes, do not lead to unacceptable large corrections to Newton's law in the four-dimensional theory. To evaluate them, note that a discrete eigenfunction of the Schr\"odinger operator with energy $m^2$ acts in four-dimensions like a field of mass $m$ and consequently, contributes with a Yukawa-like correction to the four-dimensional gravitational potential between two masses $M_1$ and $M_2$ on our brane at $z = 0$ as \cite{Csaki-UniversalAspects}
\begin{equation}
\label{Correction-Newton-Law}
\mathcal{U}(r) \approx G_N\frac{M_1 M_2}{r} + {M_{*}}^{-3}\frac{M_1 M_2 \e^{-mr}}{r}{\psi_m}^2(0),
\end{equation}
where the wave function is normalized to $\int dz \, {\psi_m}^2(0) = 1$ and $M_{*}$ is the fundamental Planck scale in $5D$. As long as $m$ increases, the correction becomes exponentially smaller.


In the Randall-Sundrum model, the static potential generated by the exchange of the zero-mode and continuum Kaluza-Klein mode propagator is \cite{RS2}
\begin{equation}
\label{CorrectionNewtonsLaw-RS}
\begin{split}
\mathcal{U}(r) \, \approx \, & G_N \frac{M_1 M_2}{r} + \int_0^{\infty} \frac{dm}{k} G_N \frac{M_1 M_2\e^{-mr}}{r}\frac{m}{k}\\
                       & = G_N\frac{M_1M_2}{r}\left(1 +\frac{1}{k^2r^2} \right).   
\end{split}
\end{equation}
The leading term due to the bound state (zero-energy mode) is the usual Newtonian potential. Furthermore, the KK modes generate an highly suppressed correction term. Such phenomena could be experimented in high-energy processes. 

It is important to note that the quantities necessary to compute the correction in the gravitational potential are the masses of the Kaluza-Klein tower and the values of the corresponding eigenfunctions at the origin. However, in general, there are not analytical solutions, and hence,  numerical approaches are desirable. In the section \ref{Sec_Results}, we will apply appropriate numerical methods to obtain such quantities in the asymmetrical hybrid brane model and in the compact brane, presented in the following two sections.


\section{Model I: The asymmetric hybrid brane}
\label{Model 1}

\begin{figure}[t]
 \centering
    \includegraphics[width=0.6\textwidth]{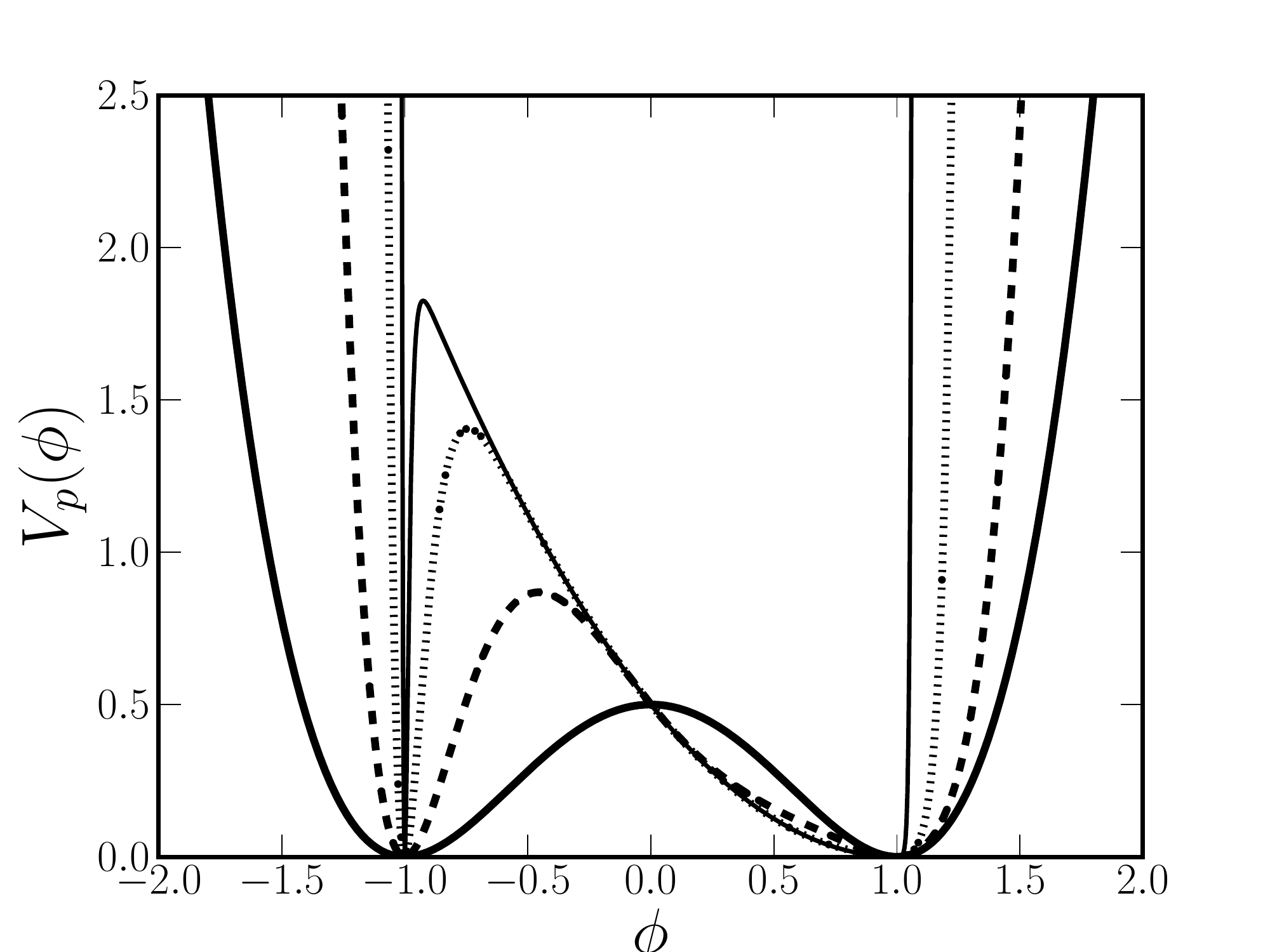}
 \caption{Potential $V_p(\phi)$ for $p = 1$ (thick line), $p = 3$ (dashed line), $p = 11$ (dotted line) and $p = 63$ (thin line). The potential function acquires a compact shape in the $\phi-$negative sector.}
 \label{Fig_Vp}
\end{figure}

The following potential \cite{AsymmetricHybridBrane}
\begin{equation}
\label{V-AsymmetricHybrid}
V_p(\phi) = \frac{1}{2}(1-\phi)^2(1+\phi^p)^2,
\end{equation}
leads to the superpotential
\begin{equation}
\label{W-AsymmetricHybrid}
W_p(\phi) = \phi - \frac{\phi^2}{2} + \frac{\phi^{1+p}}{1+p} - \frac{\phi^{2+p}}{2+p},
\end{equation}
where $p$ is an odd integer. Note that the ordinary $\phi^4$ model is recovered for $p = 1$. The potential $V_p(\phi)$ is depicted in Fig. \ref{Fig_Vp} for different values of $p$. The parameter $p$ induces a special asymmetry: differently from the symmetric hybrid brane developed in the Ref. \cite{Bazeia-FromKinksToCompactons}, the potential acquires a compact shape only in the $\phi-$negative sector. In general, the thick brane scenarios are symmetric, since the source scalar field presents a $Z_2$ symmetry. However, one can also consider an asymmetric brane which is of current interest in brane cosmology, whose contribution lies with the cosmic acceleration \cite{Padilla, FriedmanBrane}. Asymmetric branes also lead to a possible approach to the hierarchy problem \cite{AsymBlochBrane}. The asymmetry of a brane may also be thought as a consequence of different cosmological constants on each sector of the defect \cite{Ahmed-AsymThickBrane, Bazeia-AsymThickBrane}.

\begin{figure}[t]
\subfigure[]{
\includegraphics[scale=0.37]{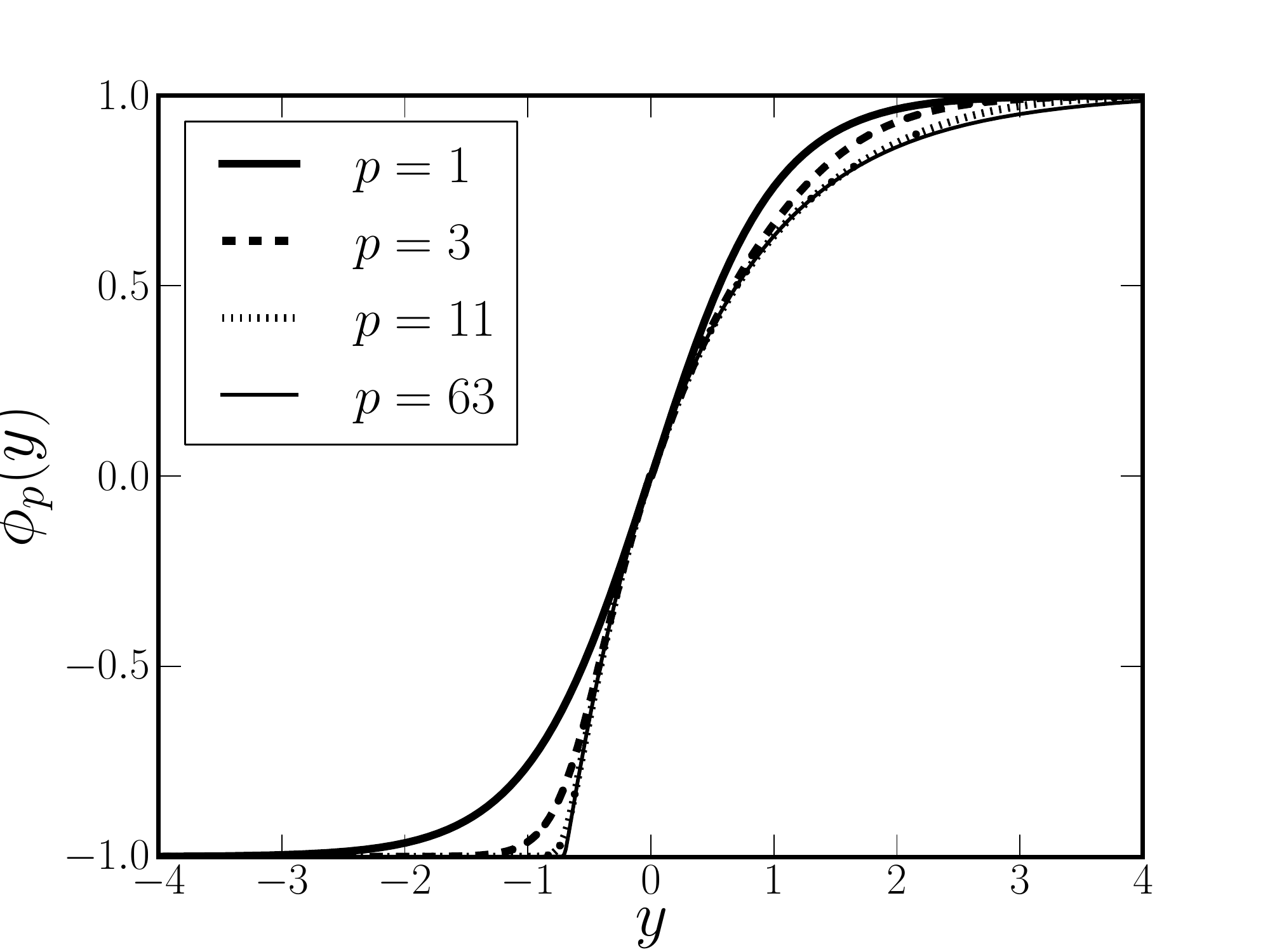}
\label{Fig_phi-p}}
\subfigure[]{
\includegraphics[scale=0.37]{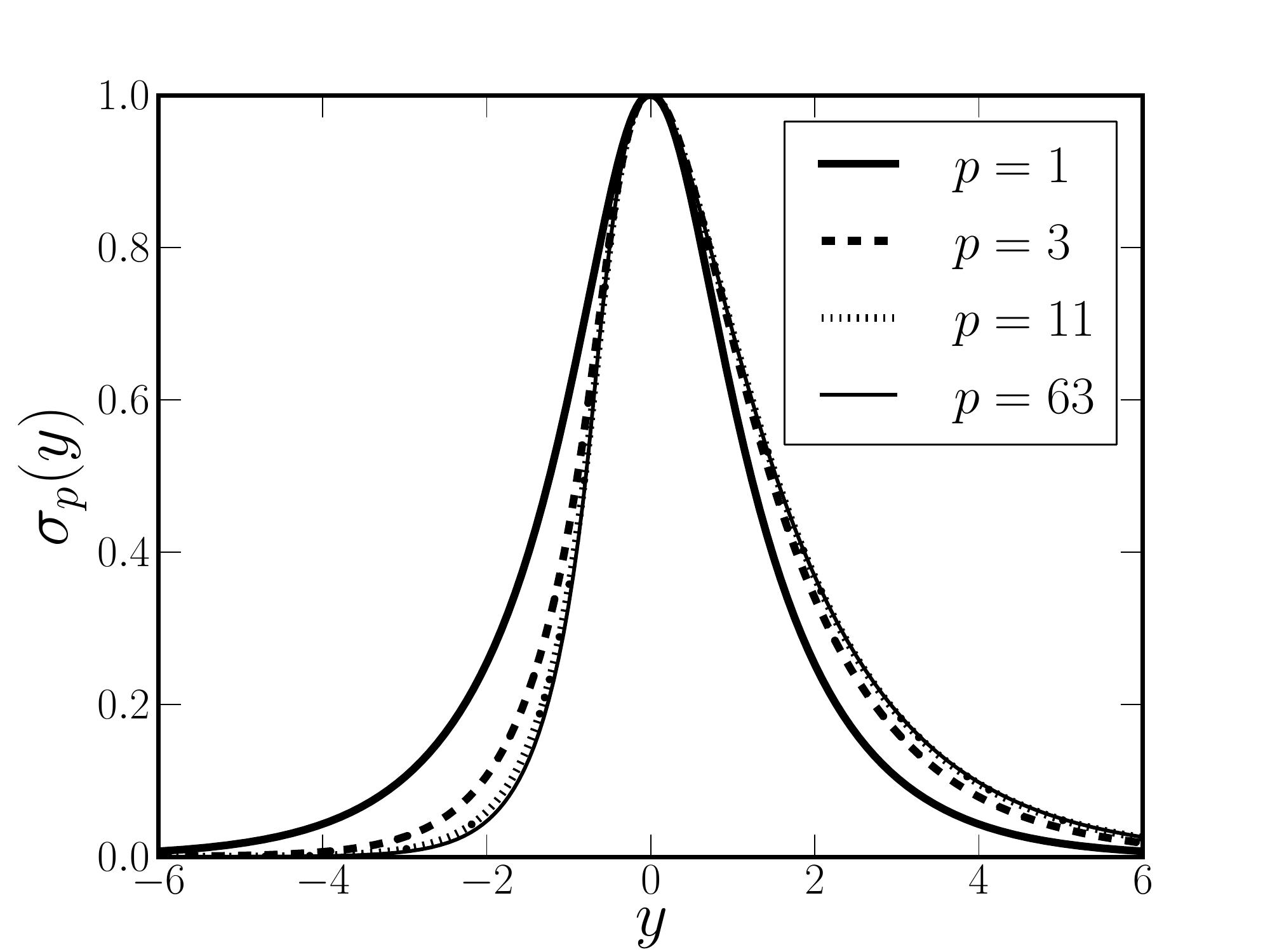}
\label{Fig_sigma-p}}
\caption{Scalar field solution $\phi_p(y)$ in (a) and warp factor $\sigma_p(y)$ in (b) for different values of $p$. The asymmetric hybrid brane is engendered by a half-compacton defect which causes an asymmetry in the warp factor.}
\label{Fig_phi-A-p}
\end{figure}

The field solution and the warp factor are obtained from the Eq. \eqref{FirstOrder_EqMotion}. Due to the complexity form of the superpotential, a numerical treatment was needed. To this aim, we solved the field equation using simple fourth-order Runge-Kutta algorithm with the boundary condition $\phi_p(0) = 0$ on each sector. For the warp factor equation, we used the shooting method with the condition $A_p(0) = 0$. We plot the approximated solutions in the Figs. \ref{Fig_phi-p} and \ref{Fig_sigma-p}. Note that the asymmetric hybrid brane is engendered by a \textit{half-compacton} defect. The hybrid profile is understood as defined in Ref. \cite{Bazeia-FromKinksToCompactons}: while the usual thick branes behave as thin branes asymptotically, the hybrid brane behaves as a thin one when the extra dimension is outside the domain where the energy density is non-trivial. In the present case, when $y < -1$. 

\begin{figure}[t] 
       \begin{minipage}[t]{0.48 \linewidth}
           \includegraphics[width=\linewidth]{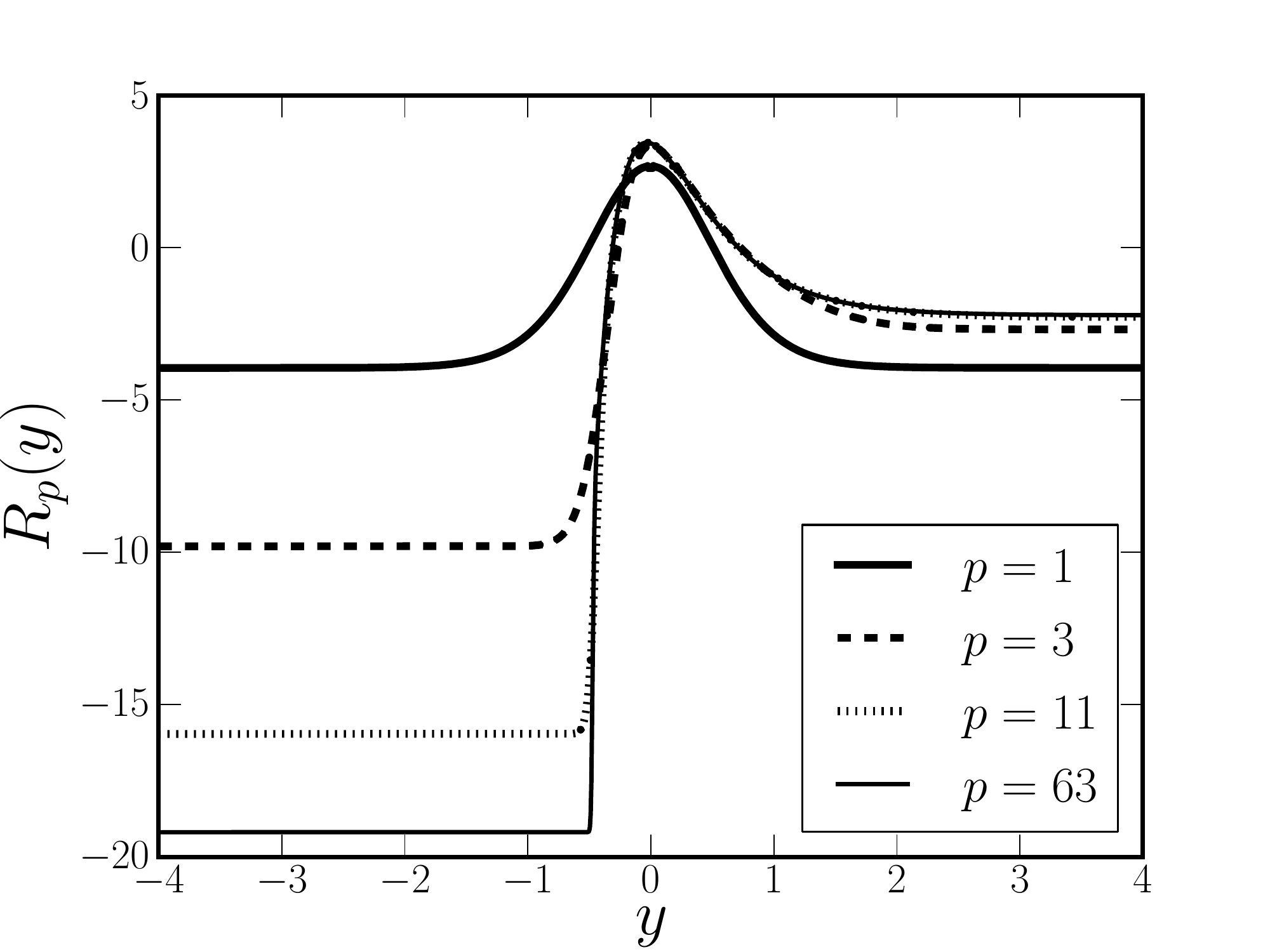}\\
           \caption{Scalar curvature for the asymmetric hybrid brane. The sudden change to a negative constant for $y\approx -0.5$ characterizes the hybrid behavior of the brane.}
          \label{Fig_R-p}
       \end{minipage}\hfill
       \begin{minipage}[t]{0.48 \linewidth}
           \includegraphics[width=\linewidth]{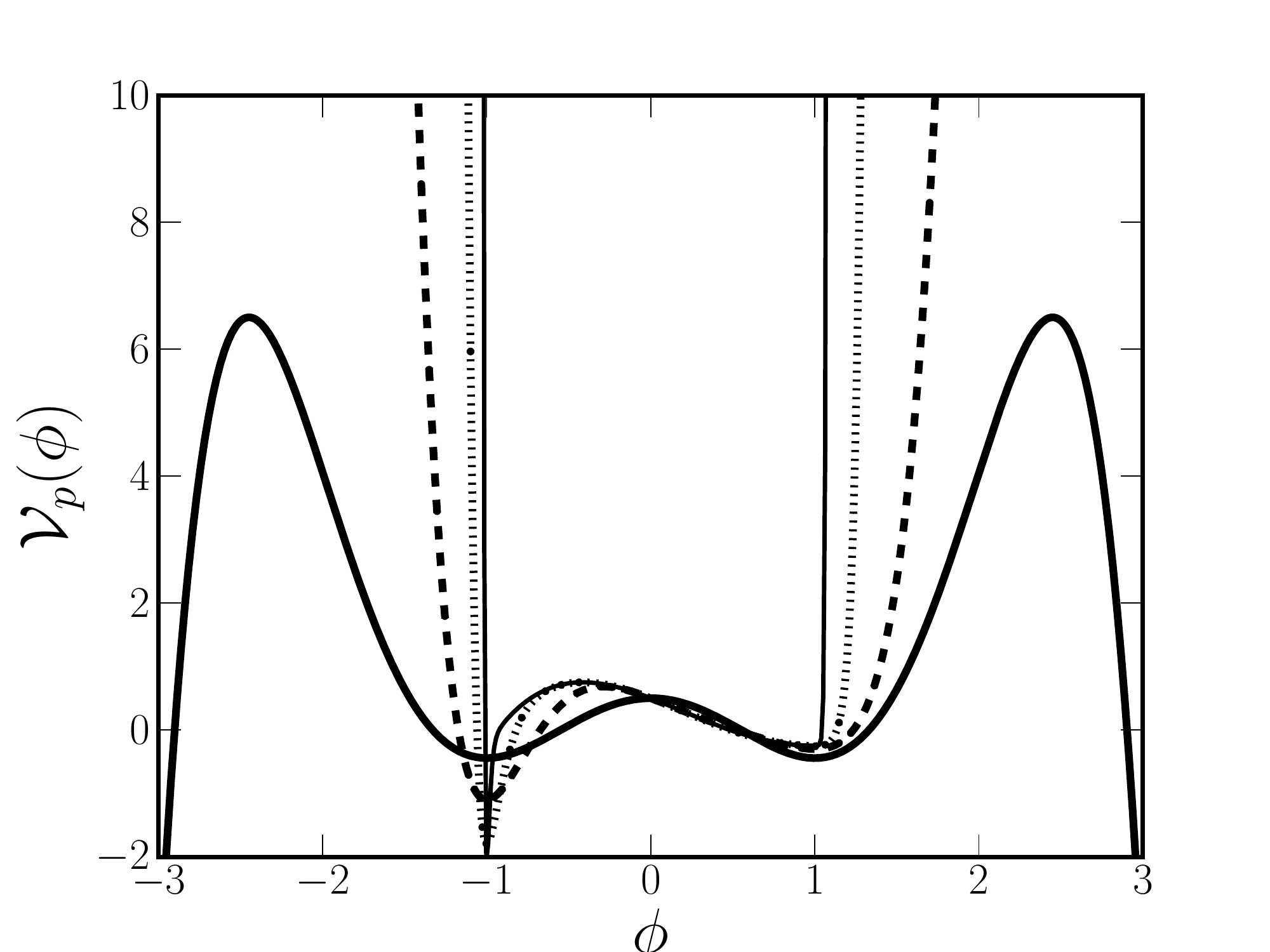}\\
           \caption{Plot of the potential in the curved space-time $\mathcal{V}_{p}$ for the same values of $p$ in Fig. \ref{Fig_R-p}. The potential acquires a limited shape in the negative part of the domain.}
           \label{Fig_Vbrana-p}
       \end{minipage}
   \end{figure}

We also calculated the scalar curvature given by Eq. \eqref{ScalarCurvature} which is plotted in Fig \ref{Fig_R-p}.  The sudden change to a negative constant characterizes the hybrid behavior of the brane. Moreover, the different asymptotic values, i.e., $|R(-\infty)| \neq |R(+\infty)|$, make clear the different values of the bulk cosmological constant on each ``side'' of the brane. We also plotted the potential in the curved space-time in Fig. \ref{Fig_Vbrana-p}. Note that it acquires a limited shape in the negative part of the domain due to the compacton piece of the scalar field. 


\begin{figure}[h]
 \centering
    \includegraphics[width=0.6\textwidth]{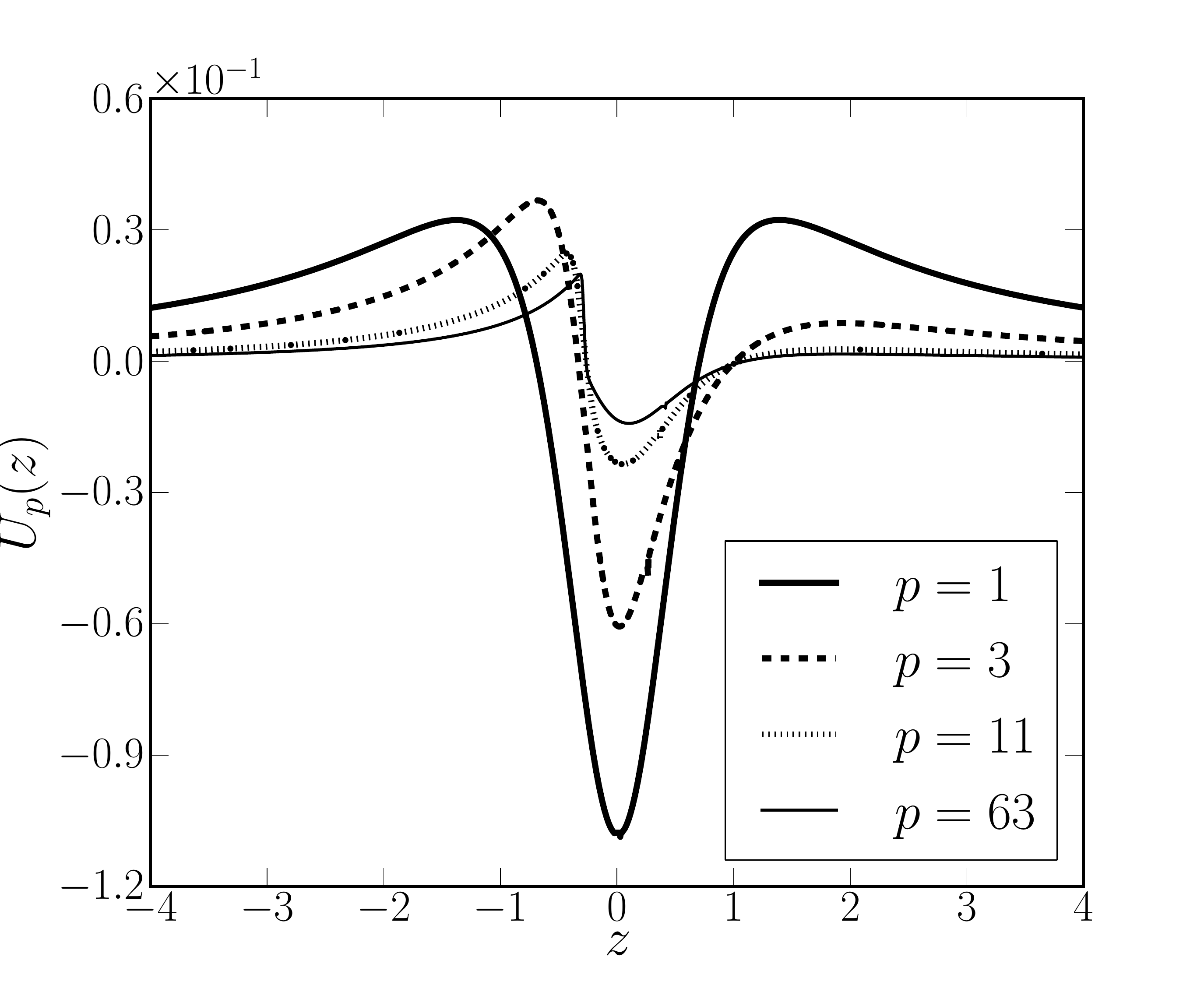}
 \caption{Analogue quantum potential for the asymmetric hybrid brane. The case for $p = 1$ coincides with the $\phi^4$ ordinary defect. As the hybrid profile evolves, the potential barrier diminishes.}
 \label{Fig_Up}
\end{figure}

Finally, we calculated the analogue quantum potential $U_p(z)$ which is plotted in Fig. \ref{Fig_Up}. The transformation of coordinates $z(y)$ was computed performing the numerical integral $z = \int \sigma^{-1/2}dy$ using the numerical approximation of the warp factor. The case for $p = 1$ coincides with the ordinary domain wall brane, as for the symmetric hybrid brane \cite{DiegoHybrid}. As the hybrid profile evolves, the potential barrier diminishes. Note the hybrid profile for negative $z$ where the potential behaves as in the thin brane case ($\approx 1/|z|^2$). It is important to mention that the small height of the potential barrier affects only small masses. 


\section{Model II: The compact brane}
\label{Model 2}

One of the newest braneworld model, proposed in the Ref. \cite{CompactBrane}, is the so-called \textit{compact brane}. Such model is engendered by the superpotential
\begin{equation}
\label{SuperPotential_Compact}
W_{\lambda}(\phi) = -\frac{1}{(1 - \lambda)} \frac{1}{\sqrt{\lambda}} \ln\left( \frac{1 - \sqrt{\lambda} \sn(\phi,\lambda)}{\dn(\phi,\lambda)} \right),
\end{equation}
where $\sn(\phi,\lambda)$ and $\dn(\phi,\lambda)$ are Jacobi elliptic functions. The Jacobi elliptic functions are a set of basic elliptic functions which are denoted by $\sn(\phi,\lambda)$, $\cn(\phi,\lambda)$ and $\dn(\phi,\lambda)$, where $\lambda$ is a parameter in the domain $[0,1]$ called elliptic modulus. 
They arise from the inversion of the elliptic integral of the first kind. 
\begin{equation}
\phi = F(u, \lambda) = \int_{0}^{u}\frac{dt}{\sqrt{1 - \lambda^2\, \sin^2 t}}.
\end{equation}
Then, the elliptic functions are given by 
\begin{equation}
\sn (\phi, \lambda) = \sin u, \hspace{1cm} \cn (\phi, \lambda) = \cos u, \hspace{0.5cm} \text{and} \hspace{0.5cm} \dn (\phi, \lambda) = \sqrt{1 - \lambda^2\, \sin^2 t}
\end{equation}

These functions satisfy the two algebraic relations:
\begin{equation}
\begin{split}
\cn^2(\phi,\lambda) + \sn^2(\phi,\lambda) & = 1,\\
\dn^2(\phi,\lambda) + \lambda^2 \sn^2(\phi,\lambda) & = 1.
\end{split}
\end{equation}
Such functions are interesting since they lead to both trigonometric and hyperbolic functions. For $\lambda = 0$, we have the usual trigonometric functions, while for $\lambda = 1$, we get the hyperbolic ones:
\begin{equation}
\begin{split}
\sn(\phi,0) = \sin \phi, & \hspace{1cm} \cn(\phi,0) = \cos \phi,\\
\sn(\phi,1) = \tanh \phi, & \hspace{1cm} \cn(\phi,1) = \dn(\phi,0) = \sech \phi.
\end{split}
\end{equation}

\begin{figure}[t]
 \centering
    \includegraphics[width=0.6\textwidth]{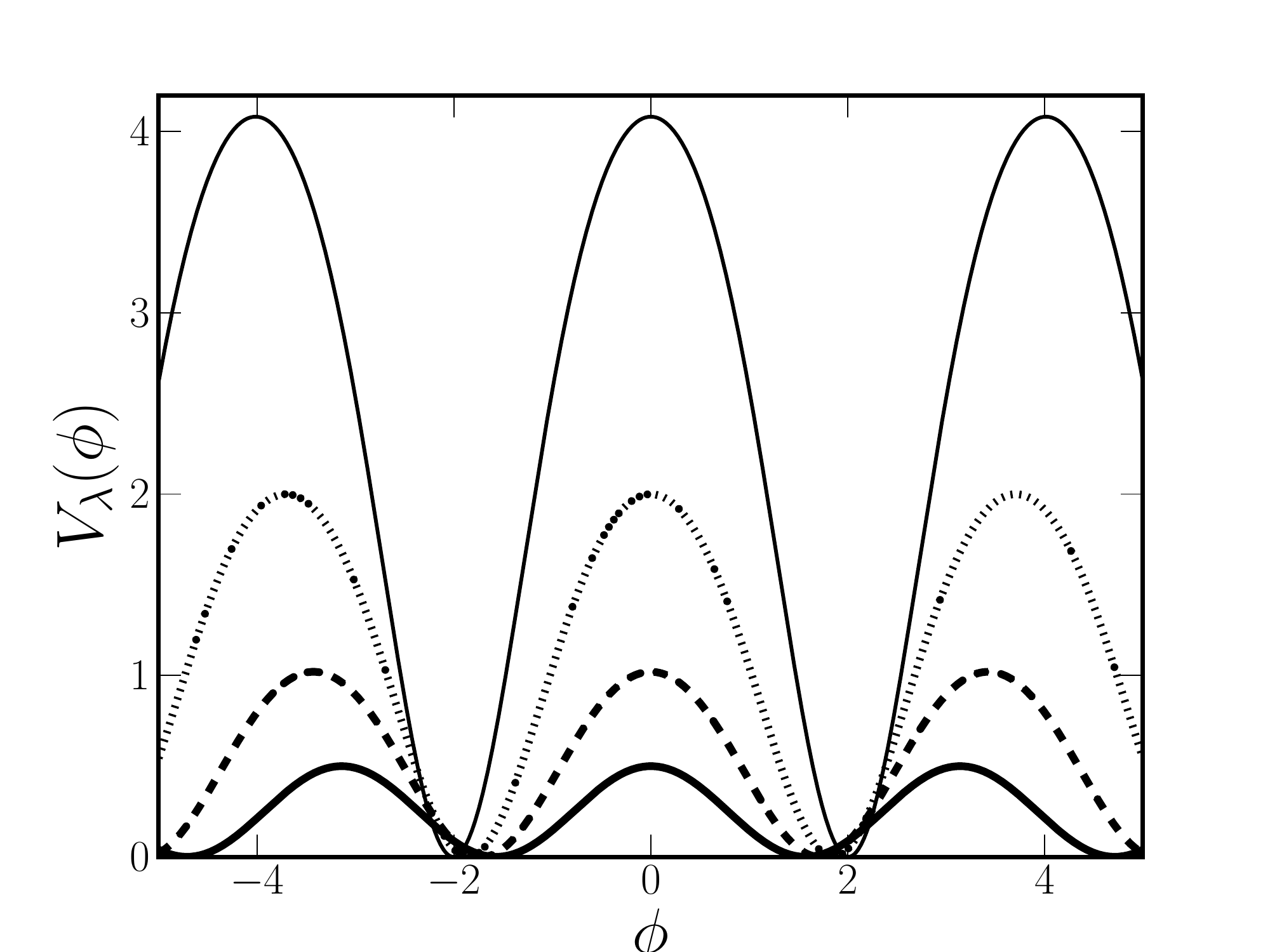}
 \caption{Potential $V_{\lambda}(\phi)$ for $\lambda=0.0$ (thick line), $\lambda=0.30$ (dashed line), $\lambda=0.50$ (dotted line) and $\lambda=0.65$ (thin line).}
 \label{Fig_V-lambda}
\end{figure}

Since the derivatives of the three basic Jacobi elliptic functions are
\begin{equation}
\begin{split}
\frac{d}{d\phi}\sn(\phi,\lambda) & = \cn(\phi,\lambda)\dn(\phi,\lambda),\\
\frac{d}{d\phi}\cn(\phi,\lambda) & = -\sn(\phi,\lambda)\dn(\phi,\lambda),\\
\frac{d}{d\phi}\dn(\phi,\lambda) & = -\lambda^2\sn(\phi,\lambda)\cn(\phi,\lambda),
\end{split}
\end{equation}
the potential of the kink-defect, given by Eq. \eqref{V(phi)=dWdphi}, becomes
\begin{equation}
\label{Potential-Compact}
V(\phi) = \frac{1}{(1-\lambda)^2}\left[\frac{1}{8}\frac{\cn^2(\phi,\lambda)}{\dn^2(\phi,\lambda)} - \frac{1}{3\lambda}\ln^2\left(\frac{1 - \sqrt{\lambda} \sn(\phi,\lambda)}{\dn(\phi,\lambda)}\right)\right],
\end{equation}
which is depicted in Fig. \ref{Fig_V-lambda} for different values of $\lambda$. This potential changes significantly with the variation of the parameter $\lambda$. Furthermore, its minima also changes. Note that for $\lambda = 0.0$, the sine-Gordon potential is recovered. 

The equation for the scalar field reads
\begin{equation}
\phi_{\lambda}^{\prime}(y) = \frac{dW_{\lambda}}{d\phi} = \frac{1}{1-\lambda}\frac{\cn(\phi,\lambda)}{\dn(\phi,\lambda)},	
\end{equation}
which has the analytical solution
\begin{equation}
\phi_\lambda(y)  = \sn^{-1}\left[\tanh\left(\frac{y}{2(1-\lambda)}\right), \lambda \right].
\label{field-equation_compact}
\end{equation}

Note that the use of the Jacobi elliptic functions in the definition of the superpotential \eqref{SuperPotential_Compact}, as proposed in the Ref. \cite{CompactBrane}, leads to a braneworld scenario whose tension is related to the elliptic modulus $\lambda$.

\begin{figure}[h]
\subfigure[]{
\includegraphics[scale=0.37]{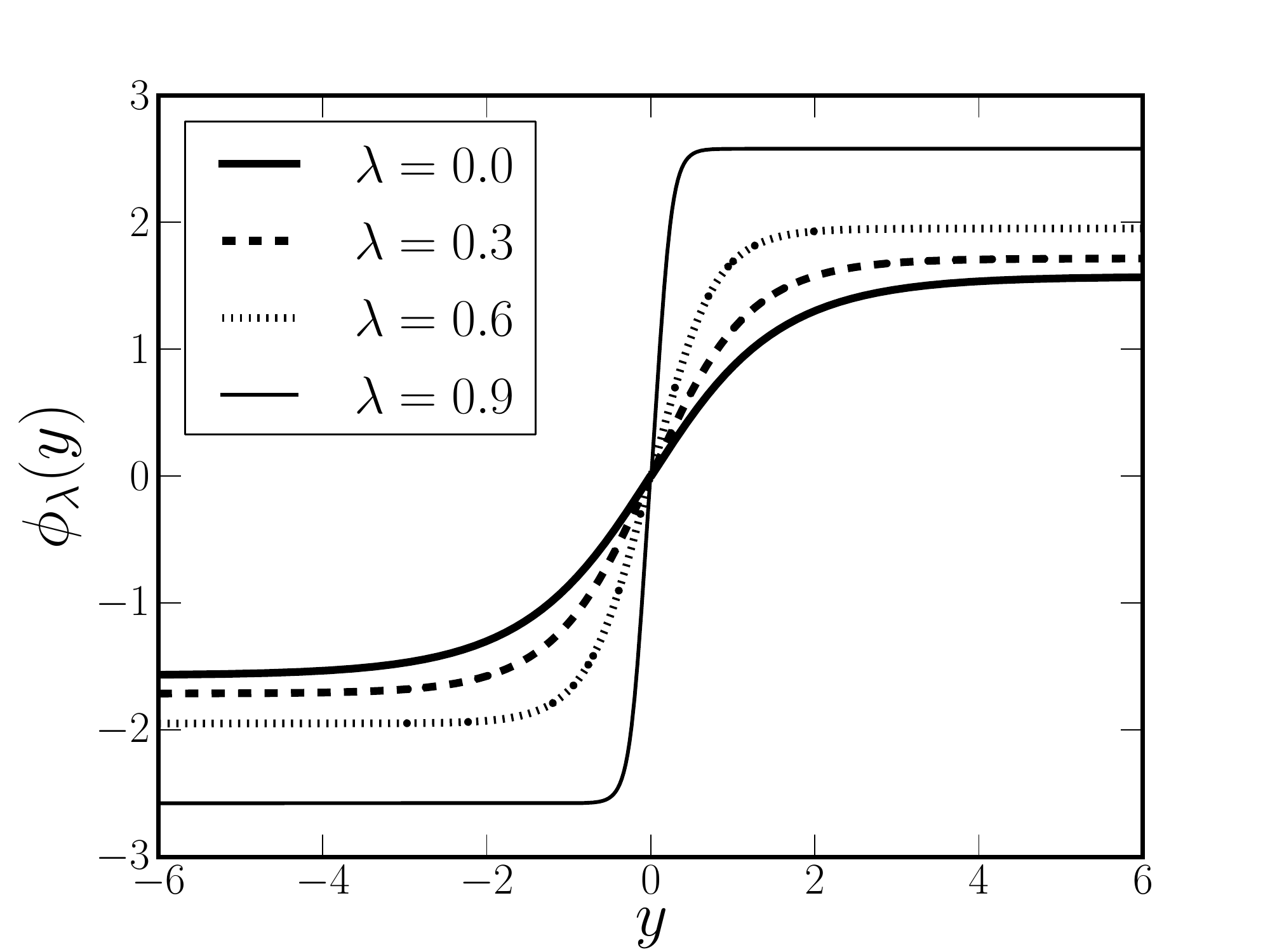}
\label{Fig_phi-lambda}}
\subfigure[]{
\includegraphics[scale=0.37]{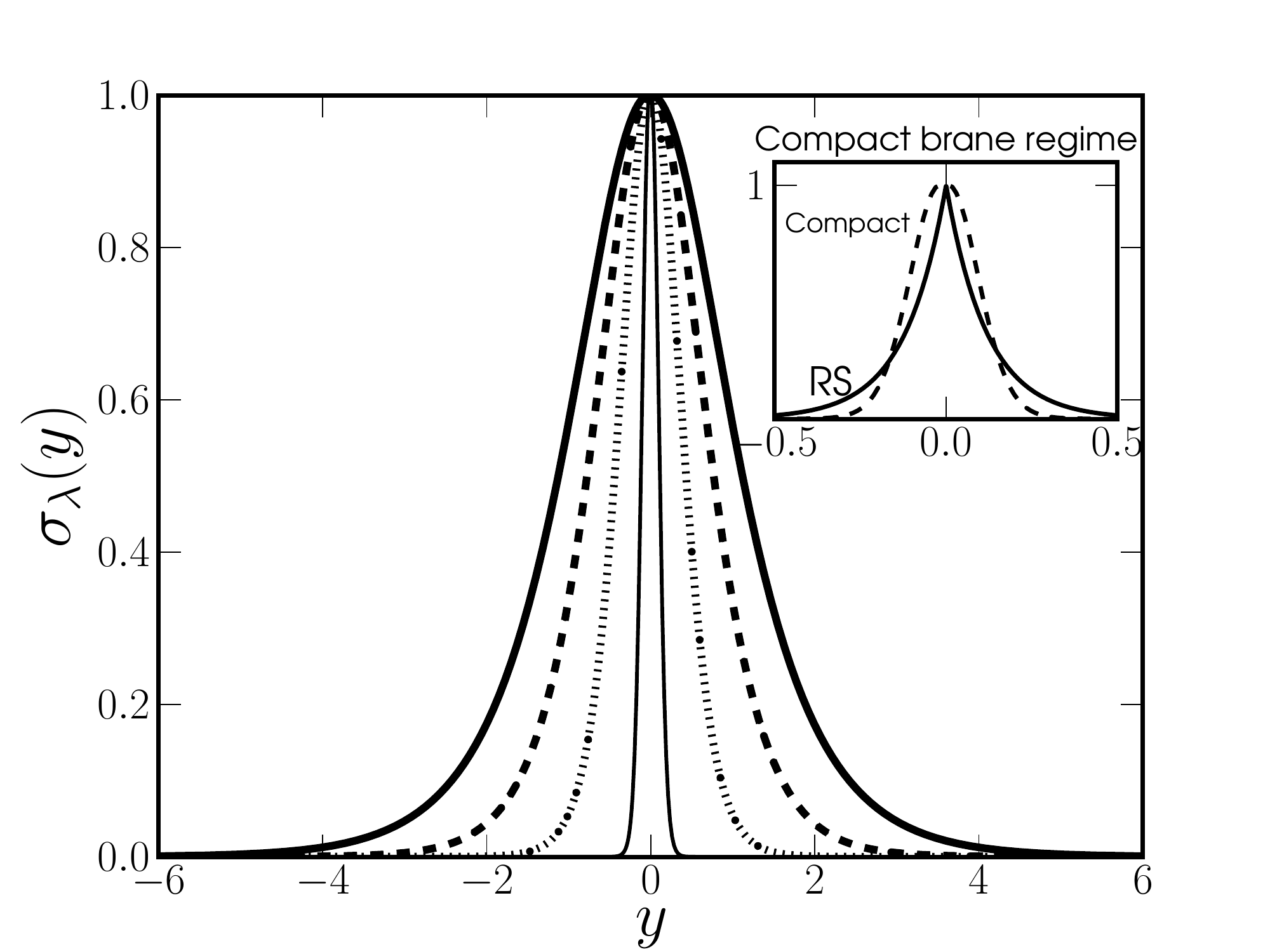}
\label{Fig_sigma-lambda}}
\caption{Scalar field solution $\phi_{\lambda}(y)$ in (a) for different values of the parameter $\lambda$. Warp factor $\sigma_{\lambda}(y)$ in (b) for the same values of $\lambda$. As $\lambda$ increases, the warp factor becomes more and more compact. In the sub-figure, we plot the warp factor in the compact regime $(\lambda = 0.95)$ and the Randall-Sundrum model with $k = 30.5$.}
\label{Fig_phi-A-lambda}
\end{figure}

We obtained numerically the solution for the warp factor given by the Eq. \eqref{FirstOrder_EqMotion} using the same methods employed in the \textit{asymmetric hybrid brane} of the previous section. We plot in Figs \ref{Fig_phi-lambda} and \ref{Fig_sigma-lambda} the scalar field and warp factor, respectively. The field solution becomes compact (do not make confusing with the \textit{compacton} defect discussed in the previous section) and the asymptotic values increases with $\lambda$, which agrees with the displacement of the minima of the potential function $V_{\lambda}(\phi)$ displayed in the Fig. \ref{Fig_V-lambda}. Note that when $\lambda\rightarrow 1$, the width of the warp factor decreases significantly making the extra dimension compact. In the sub-graph of Fig. \ref{Fig_sigma-lambda}, we plot together with the compact brane case, the warp factor of the Randall-Sundrum model with $k = 30.5$, i.e., $\sigma_{RS} = \e^{-70.0|y|}$. Hence, it may be concluded that the compact brane corresponds to a brane with very high curvature. This issue prevents the sustenance of massive states, as we will show in the next section.

\begin{figure}[t] 
       \begin{minipage}[t]{0.49 \linewidth}
           \includegraphics[width=\linewidth]{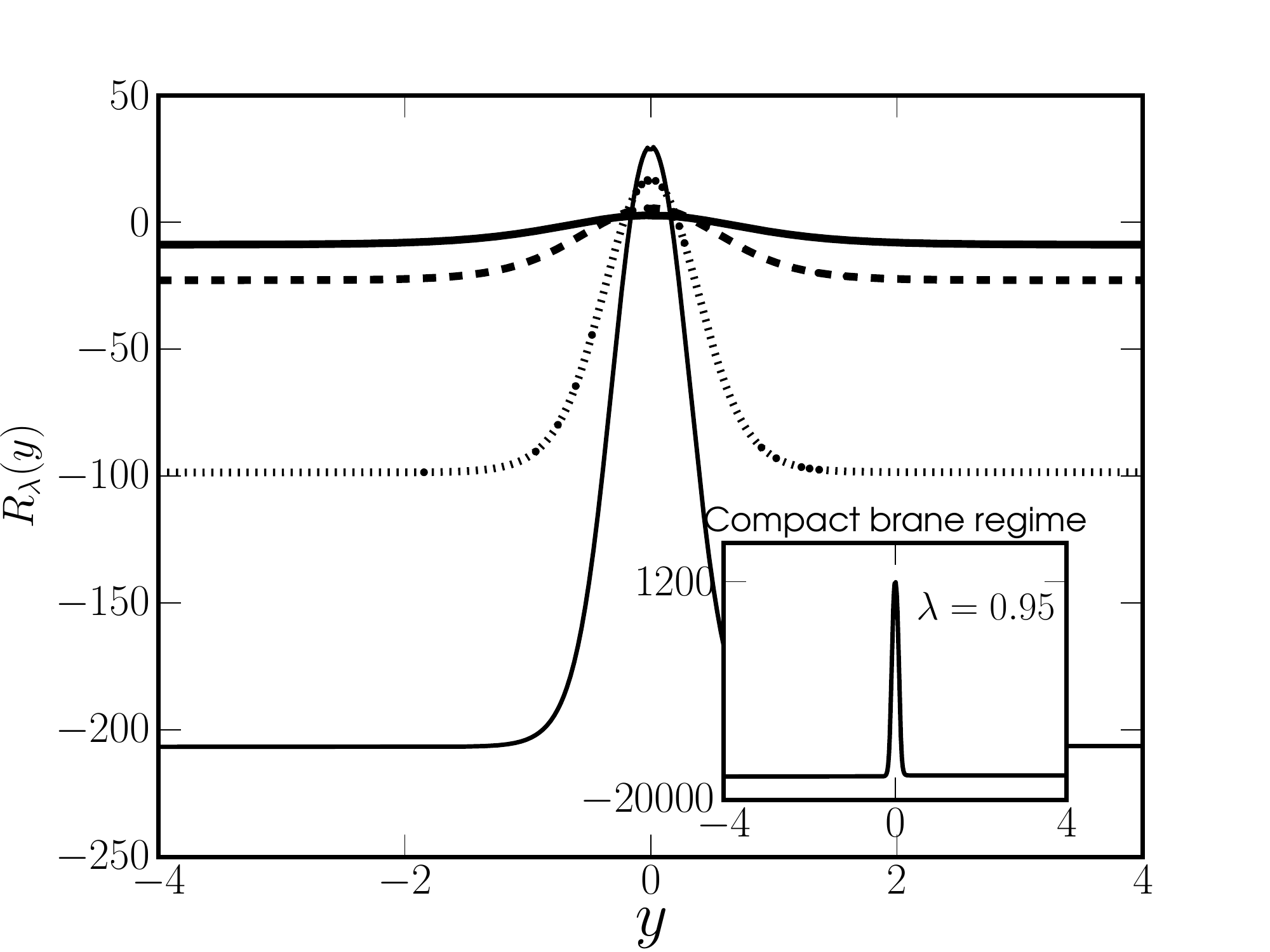}\\
           \caption{Scalar curvature for the compact brane model for the same parameter values of the Fig. \ref{Fig_phi-A-lambda}. As the parameter $\lambda$ increases, the brane becomes more and more compressed around the origin. The sub-figure shows the compact regime.}
          \label{Fig_R-lambda}
       \end{minipage}\hfill
       \begin{minipage}[t]{0.49 \linewidth}
           \includegraphics[width=\linewidth]{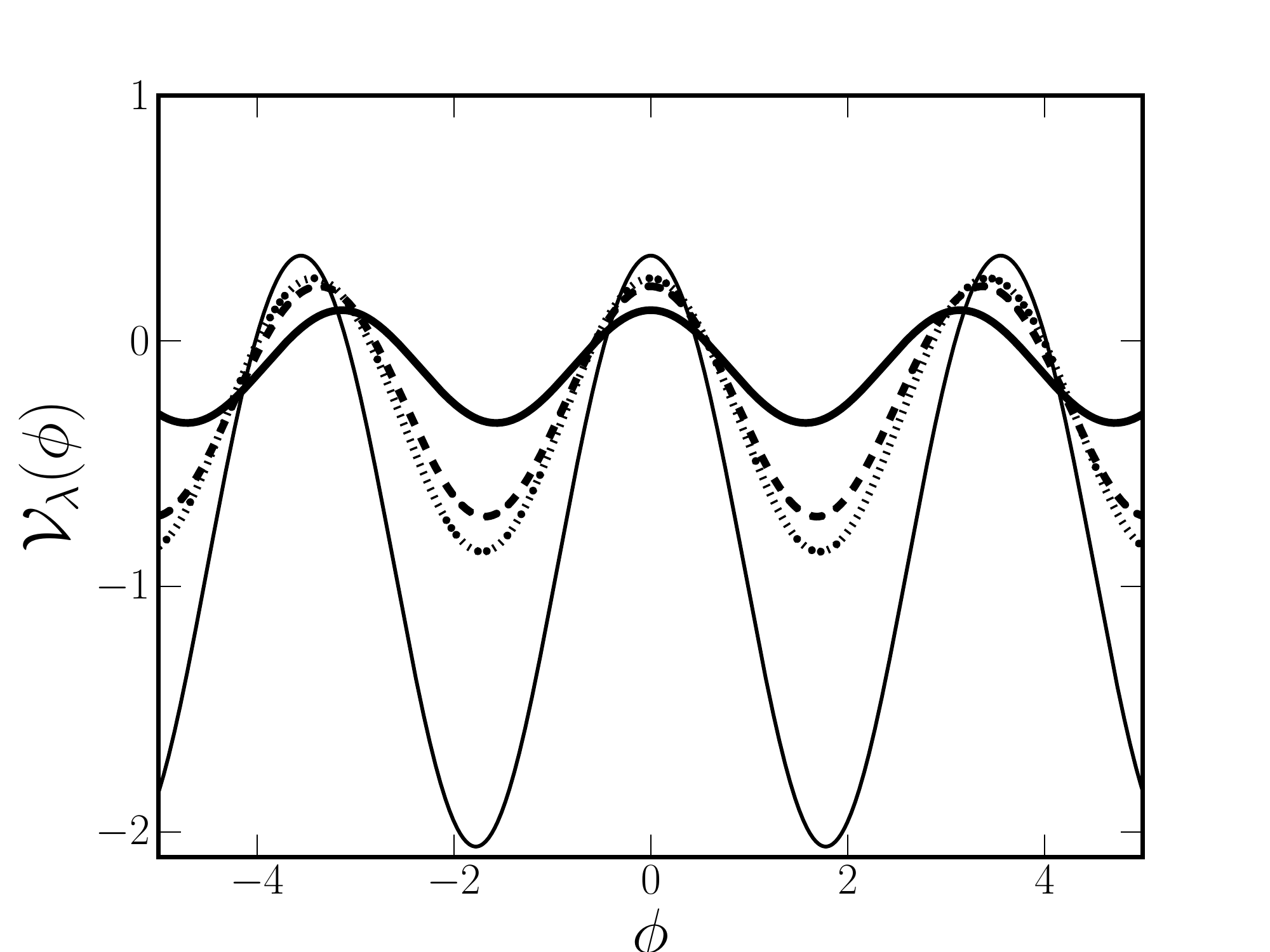}\\
           \caption{Plot of the potential in curved space-time $\mathcal{V}_{\lambda}(\phi)$ for the same parameter values of the Fig. \ref{Fig_phi-A-lambda}. For $\lambda\rightarrow 0$, the potential coincides with the sine-Gordon model.}
           \label{Fig_Vbrana-lambda}
       \end{minipage}
   \end{figure}

To reinforce the compact profile of the brane, we computed numerically the scalar curvature given by Eq. \eqref{ScalarCurvature}. We plot the numerical solution in the Fig. \ref{Fig_R-lambda}. As the parameter $\lambda$ increases, the brane becomes more and more compressed around the origin. Note also that the constant asymptotic values of the scalar curvature become very negatively large as the brane tends to the compact regime.

The potential in the curved space-time, given by the Eq. \eqref{Potential-Brane}, is obtained analytically from the superpotential \eqref{SuperPotential_Compact} as: 
\begin{equation}
\mathcal{V}_\lambda(\phi) = \frac{1}{(1-\lambda)^2}\left[ \frac{1}{8}\frac{\cn^2(\phi,\lambda)}{\dn^2(\phi,\lambda)} - \frac{1}{3\lambda}\ln^2\left( \frac{1-\sqrt{\lambda}\sn(\phi,\lambda)}{\dn(\phi,\lambda)} \right) \right].
\end{equation}
The use of the Jacobi elliptic functions in the definition of the superpotential is very interesting since it interpolates continuously from the sine-Gordon model to the compact brane model. Note that the limit $\lambda\rightarrow 0$, leads to the particular case of the sine-Gordon model studied in Ref. \cite{Gremm} and recently in the Ref. \cite{sine-GordonBrane}. See Fig. \ref{Fig_Vbrana-lambda}.


\begin{figure}[b]
 \centering
    \includegraphics[width=0.6\textwidth]{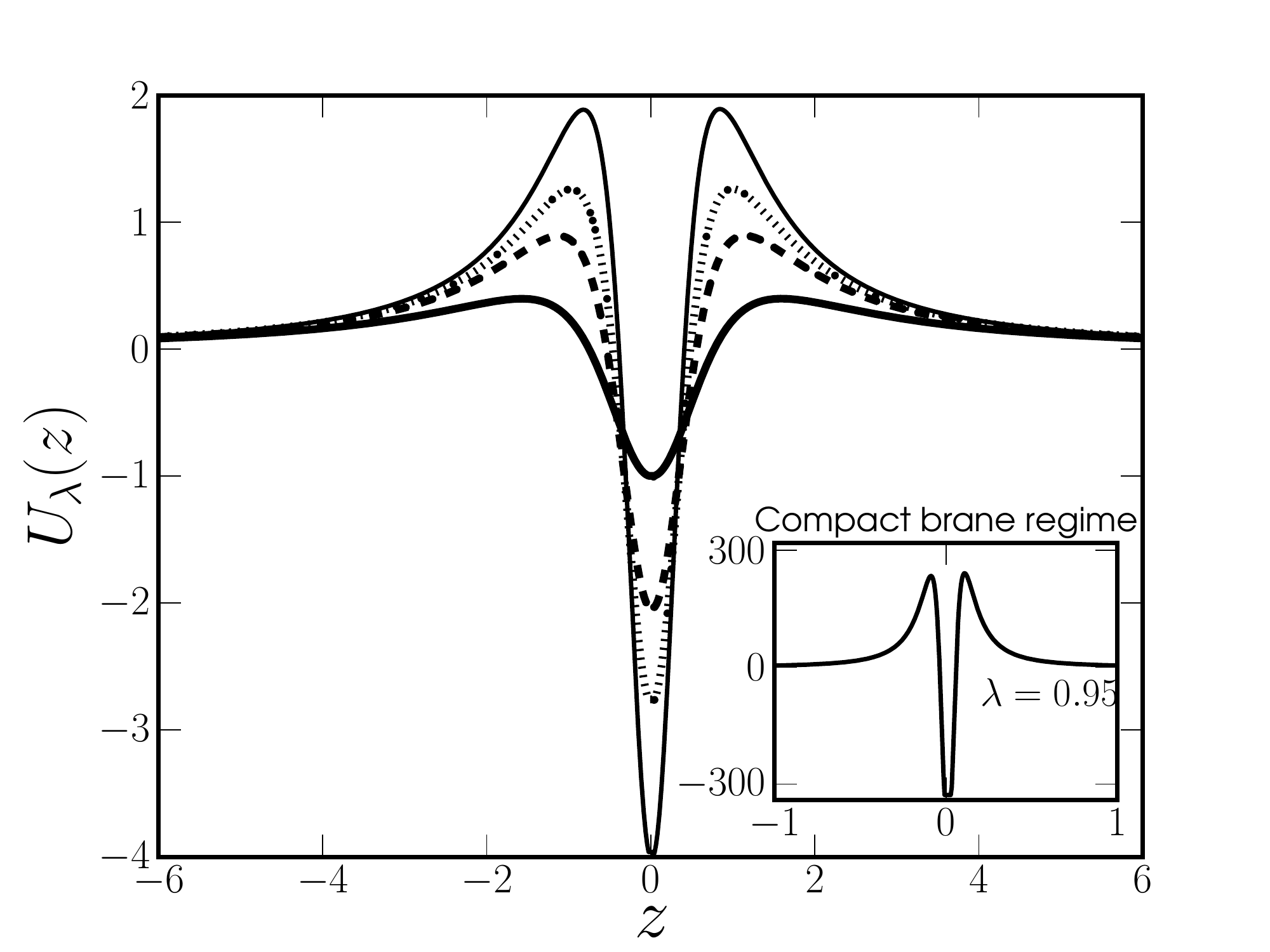}
 \caption{Analogue quantum potential $U_\lambda(\phi)$ for $\lambda = 0.0$ (thick line), $\lambda = 0.30$ (dashed line), $\lambda = 0.40$ (dotted line) and $\lambda = 0.50$ (thin line). The case for $\lambda = 0.0$ coincides with the sine-Gordon braneworld. The potential is very sensible in the parameter $\lambda$ and, as the compact profile arises, the potential barrier increases quickly. The sub-graph corresponds to the scale adjustment of the $\lambda = 0.95$ case.}
 \label{Fig_Ulambda}
\end{figure}

Finally, we computed the analogue quantum potential given by Eq. \eqref{Potencial-Schr}, which is plotted in the Fig. \ref{Fig_Ulambda} for different values of $\lambda$. The case for $\lambda = 0.0$ coincides with the sine-Gordon braneworld \cite{sine-GordonBrane}. The analogue quantum potential is very sensible with variations of the parameter $\lambda$ and its barrier increases quickly, evincing the compact profile and the high curvature of the brane. In the compact limit, which is plotted for $\lambda = 0.95$ in the sub-graph of Fig. \ref{Fig_Ulambda}, the barrier is three orders of magnitude higher than the sine-Gordon case. This issue hinders the presence of bound states, since the height of the barrier is overly high and the width of the well becomes extremely sharp. In the next section, we will present our numerical results from the computation of the Kaluza-Klein tower of states. We will show the absence of massive states in the compact brane regime.

\section{Numerical results}
\label{Sec_Results}

We are interested in computing the corrections in the Newton's law of gravitation due to the gravitational massive modes in the thick braneworld models proposed in the Refs. \cite{AsymmetricHybridBrane, CompactBrane}. To this aim, it is necessary to attain the mass spectrum solving the Eq. \eqref{Graviton-Sturm} or Eq. \eqref{Graviton-Schr}. However, such equations have no analytical solution. Fortunately, the so-called matrix method \cite{MatrixMethod} is an efficient numerical method for solving Sturm-Liouville problems. In the braneworld context, this technique was successfully applied in codimension-$2$ models \cite{Julio, DiegoGraviton, DiegoGauge, DiegoFermion} and recently, in the symmetric hybrid brane model \cite{DiegoHybrid}.

\begin{figure}[!hb]
 \centering
    \includegraphics[width=0.6\textwidth]{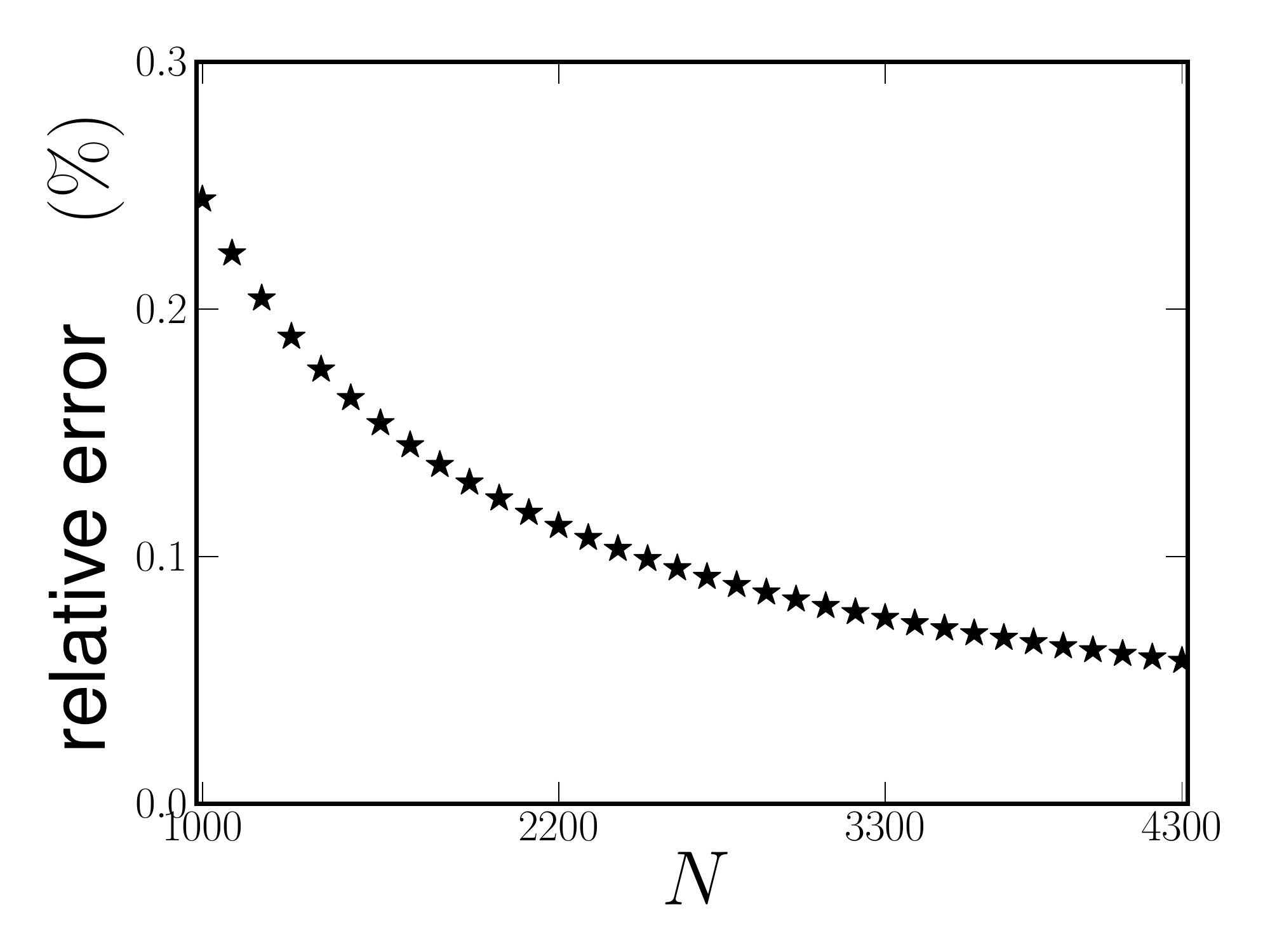}
 \caption{Relative error in the approximation  of the first mass eigenvalue of the Randall-Sundrum model in terms of the number of steps in the discretized domain.}
 \label{Fig_Erro}
\end{figure}

First of all, we applied the matrix method to the quite known Randall-Sundrum model to test our routines. To get an efficient approximation for the first twenty eigenvalues, we have discretized the differential equation \eqref{Graviton-Sturm-ExtraCoord} in $N = 4 \, 300$ steps in the domain $[-6.0, \, 6.0]$, where the warp factor is sufficiently close to zero near the boundary of the box. In this uniform grid, the stepsize is $0.002790$. We used centered finite differences approximation for the derivatives in Eq. \eqref{Graviton-Sturm-ExtraCoord} with second order truncation error. Hence, an eigenvalue problem for a $4 \, 300 \times 4 \, 300$ tridiagonal matrix had to be solved. Moreover, since only the value of the wave function at the origin is relevant to compute the correction in the Newton's law, we relinquished the unphysical boundary condition \cite{Gremm}, untying its value at $z = 0$, and adopted homogeneous Neumann boundary conditions at the edges of the box. Note that such conditions also conduct to the spectrum given by Eq. \eqref{RS-Spectrum}. Moreover, the homogeneous Neumann boundary conditions at $y = \pm L$ is in accordance with the fact that the solutions $\psi(z)$ must behave as plane waves far from the brane. In the table \ref{Tabela-Espectro}, we present the exact solution \eqref{RS-Spectrum} and the numerical approximation for $k = 0.7$ and $L = 6.0$. Note that the relative error is predominantly less than $0.06\%$. Furthermore, the approximation deteriorates with $n$ after $n = 14$. This is a natural attribute of the matrix method \cite{MatrixMethod}. However, this issue does not interferes in our analysis, since only the small masses are of physical interest \cite{RS2, Metastable, Csaki-UniversalAspects, Csaki-Quasilocalization}. To verify the stability of the numerical method, we computed the first eigenvalue for several values of $N$. We present the relative error in the Fig. \ref{Fig_Erro}. Note that the error reduces monotonically as the step-size diminishes.

\begin{figure}[t]
 \centering
    \includegraphics[width=0.6\textwidth]{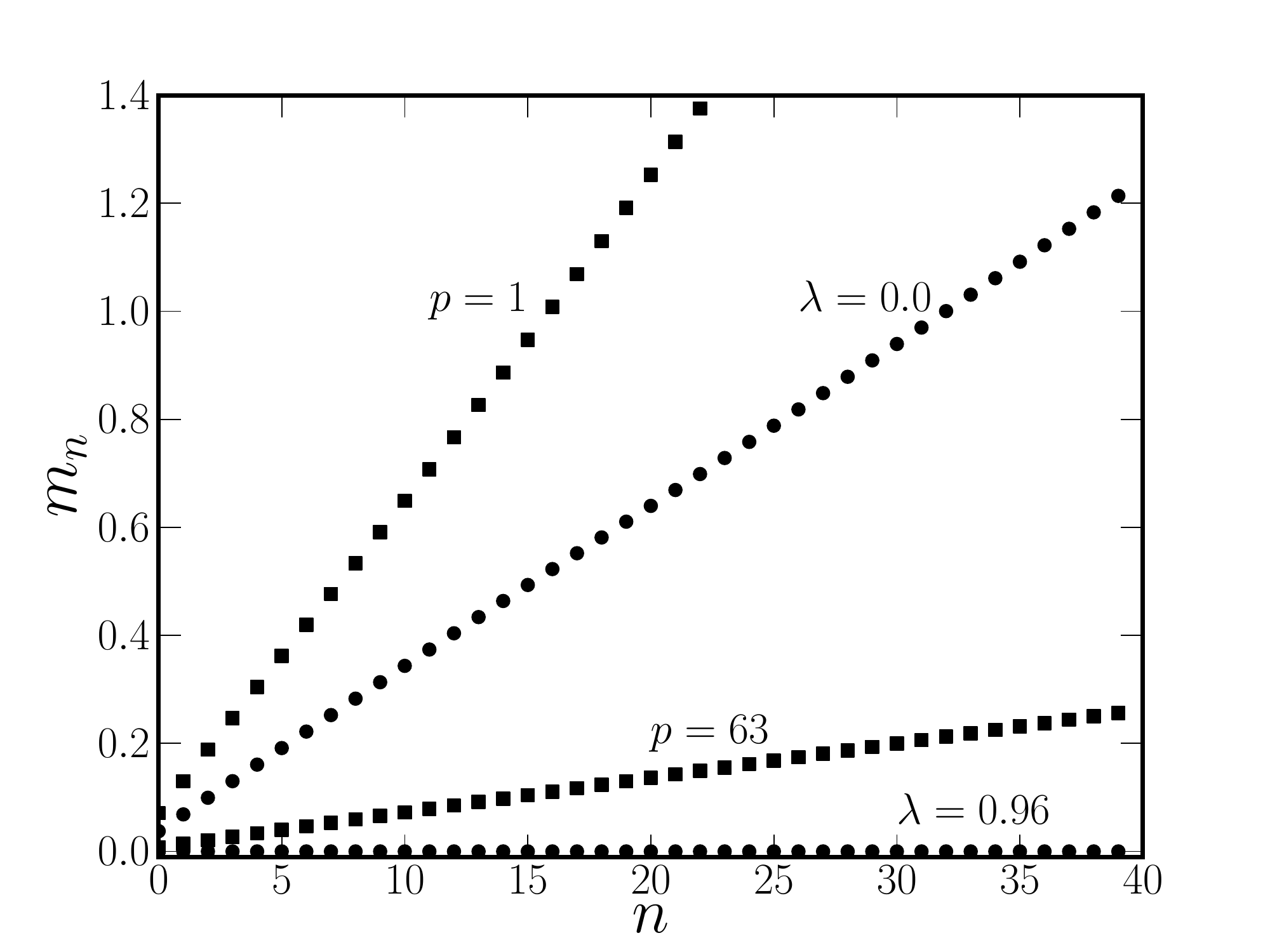}
 \caption{Graviton mass spectrum for the asymmetric hybrid brane (squares) and compact brane (circles). The spectrum is model-dependent and the whole sets decrease with the parameters. For the compact brane ($\lambda = 0.96$), the spectrum is trival.}
 \label{Fig_Spectra}
\end{figure}

After testing the efficiency of our routines, we have applied the matrix method to attain the KK spectrum for the asymmetric hybrid brane and for the compact brane. We have studied the cases $p = 1$ ($\phi^4$ model), $p = 63$ (asymmetric hybrid brane), $\lambda = 0.0$ (sine-Gordon brane) and $\lambda = 0.95$ (compact brane). We plot in Fig. \ref{Fig_Spectra} the spectra for all the mentioned cases. Note that the mass spectrum is model-dependent and the whole set decreases as the parameters (related to the brane tension) increase. In the compact brane regime, the spectrum is trivial. In fact, this is consistent with the thin brane case with very high curvature, whose first eigenvalue for $k = 30.5$ is $m_1^{RS} = 3.90\times 10^{-78}$ (see Eq. \eqref{RS-Spectrum}). Then, we may conclude that the compact brane does not have contributions to the corrections in the Newton's law. For $p = 1$ in the hybrid model (domain wall brane), the spectrum matches with our previous results reported in Ref. \cite{DiegoHybrid}. The first four eigenvalues are $m_1 = 0.0738777$, $m_2 = 0.135154$, $m_3 = 0.195893$ and $m_4 = 0.256381$. Moreover, in the hybrid regime, the spectrum becomes one order of magnitude lower than the domain wall regime. Then, this model will have a lower contribution to the correction in the Newtonian potential.

\begin{table}[t]
\centering
 \begin{tabular}{||c|c|c|c||} 
 \hline
 $n$ & $m_n = k\e^{-kL}j_n$  & Numerical Approximation & Relative error (\%) \\ [1.ex] 
 \hline\hline
 1 & 0.0402209 & 0.0402442 & 0.0578 \\ 
 2 & 0.0740640 & 0.0740264 & 0.0508 \\
 3 & 0.106790 & 0.106849 & 0.0557 \\
 4 & 0.139857 & 0.139917 & 0.0430 \\
 6 & 0.172890 & 0.172951 & 0.0352 \\  
 7 & 0.205906 & 0.205966 & 0.0293 \\  
 8 & 0.238910 & 0.238971 & 0.0256 \\ 
 9 & 0.271908 & 0.271970 & 0.0228 \\ 
 10 & 0.304901 & 0.304966 & 0.0211 \\ 
 11 & 0.337892 & 0.337959 & 0.0198 \\
 12 & 0.370879 & 0.370951 & 0.0192 \\
 13 & 0.403866 & 0.403942 & 0.0188 \\
 14 & 0.436850 & 0.436933 & 0.0190 \\
 15 & 0.469834 & 0.469925 & 0.0194 \\
 16 & 0.502817 & 0.502917 & 0.0199 \\ [1ex] 
 \hline
 \end{tabular}
 \caption{Mass eigenvalues in the Randall-Sundrum model and its numerical approximation.}
 \label{Tabela-Espectro}
\end{table}


%
%
%
%
%
%
%
%
%


\begin{figure}[t]
 \centering
    \includegraphics[width=0.6\textwidth]{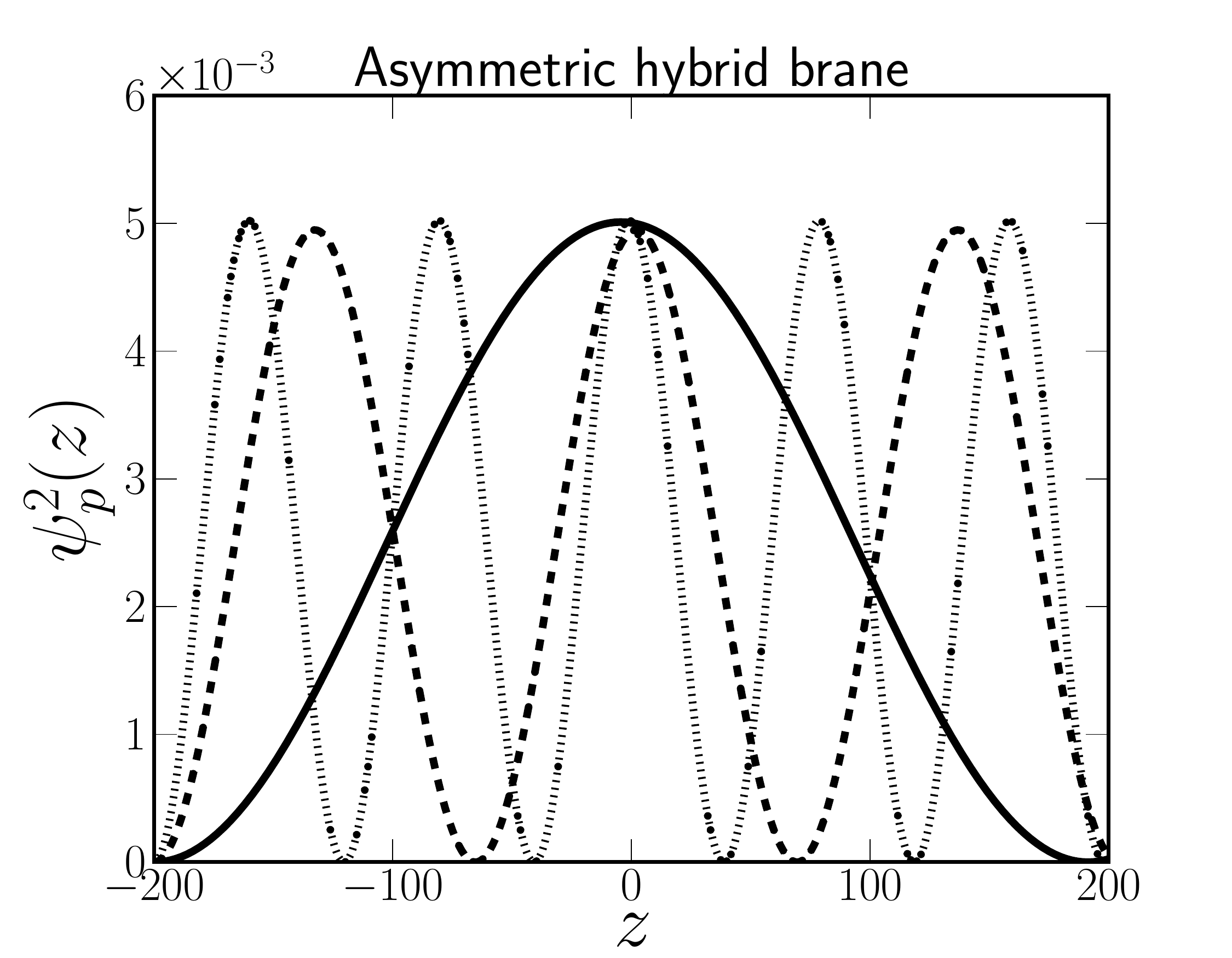}
 \caption{Normalized squared odd wave functions for the asymmetric hybrid brane ($p = 63$). The mass eigenvalues from Fig. \ref{Fig_Spectra} are $m_1 = 0.00804991$ (thick line), $m_3 = 0.0213199$ (dashed line) and $m_5 = 0.0344917$ (dotted line).}
 \label{Fig_Wavefunction-Asym}
\end{figure}

\begin{figure}[!h]
 \centering
    \includegraphics[width=0.6\textwidth]{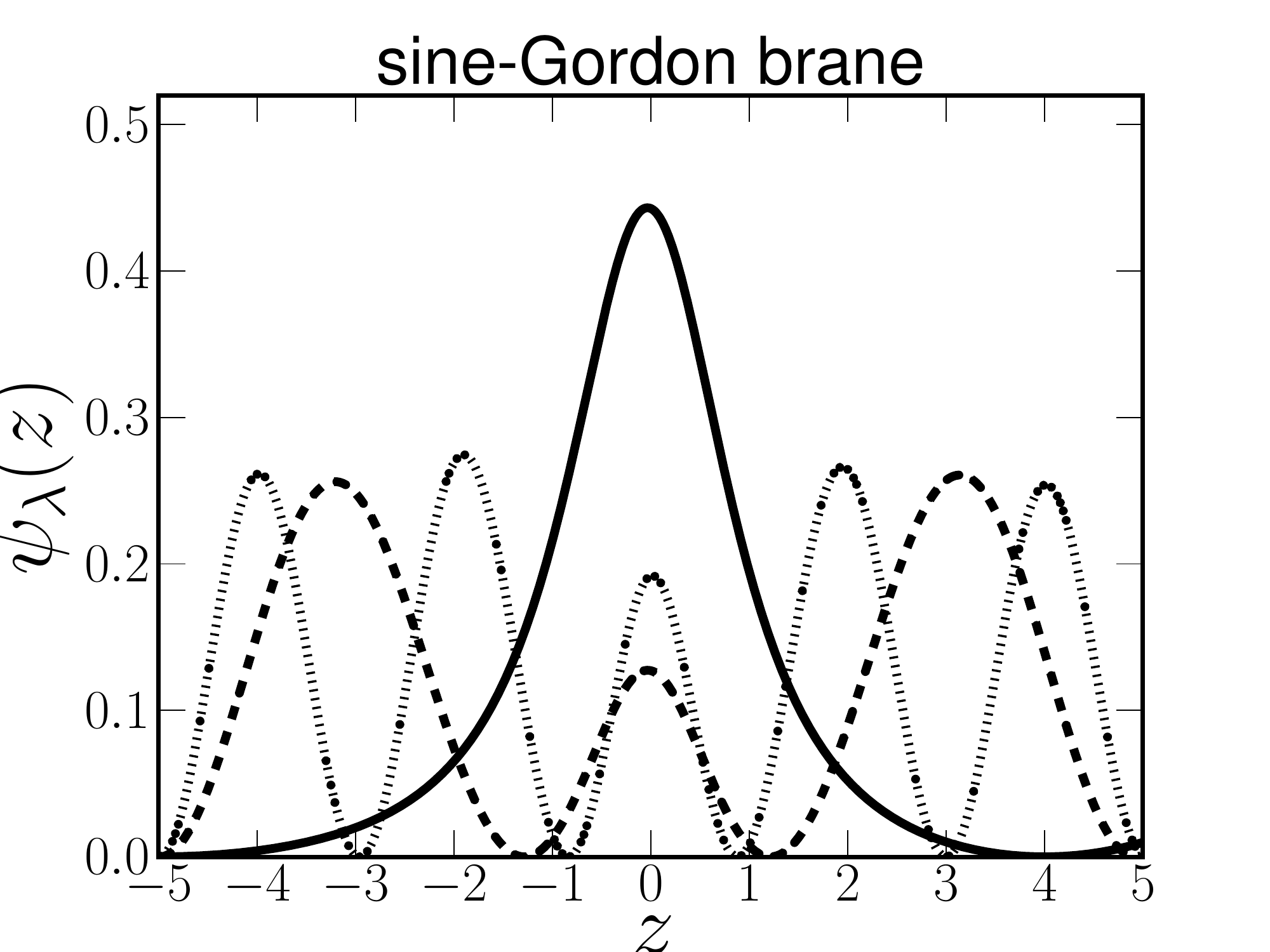}
 \caption{Normalized squared odd wave functions for the sine-Gordon brane ($\lambda = 0.0$). All even wave functions vanish in $z = 0$, so we omitted here. The mass eigenvalues from the Fig. \ref{Fig_Spectra} are $m_1 = 0.0725202$ (thick line), $m_3 = 0.192295$ (dashed line) and $m_5 = 0.310824$ (dotted line).}
 \label{Fig_Wavefunction-sineGordon}
\end{figure}

Finally, we have solved the Schr\"{o}dinger-like equation for the cases $p = 63$ and $\lambda = 0.0$. We have forsaken the compact brane regime, since only $m = 0$ is supported. We used the well-known Numerov method \cite{Numerov} for the mass eigenvalues obtained by the matrix method. The Numerov method was largely applied in field localization on branes \cite{Numerov-Adalto, Wilami-Resonance-Deformed, CASA-Ressonancia, Numerov-Chineses01, Numerov-Hott, Numerov-Chineses02, Numerov-Wilami}. We plot the normalized solutions for the first three odd eigenfunctions in Figs. \ref{Fig_Wavefunction-Asym} and \ref{Fig_Wavefunction-sineGordon}. All the even solutions vanishes at $z = 0$, then we discarded them since they do not contribute to the correction in the Newtonian potential (see Eq. \eqref{Correction-Newton-Law}). The solutions for the asymmetric hybrid brane are purely plane waves. Since all the values at $ z = 0$ are approximately equal to $0.0050$, all the odd eigenfunctions will contribute equally for the correction in the gravitational potential. On the other hand, for the sine-Gordon brane, the contributions of the eigenfunctions are different, thus all the odd eigenvalues must be taken into account in the sum of Eq. \eqref{Correction-Newton-Law}. Note further that the first massive eigenstate has the highest contribution. Therefore, possible effects of the Kaluza-Klein tower in high energy collision may be arising from the first KK mass. 

\begin{figure}[h]
 \centering
    \includegraphics[width=0.6\textwidth]{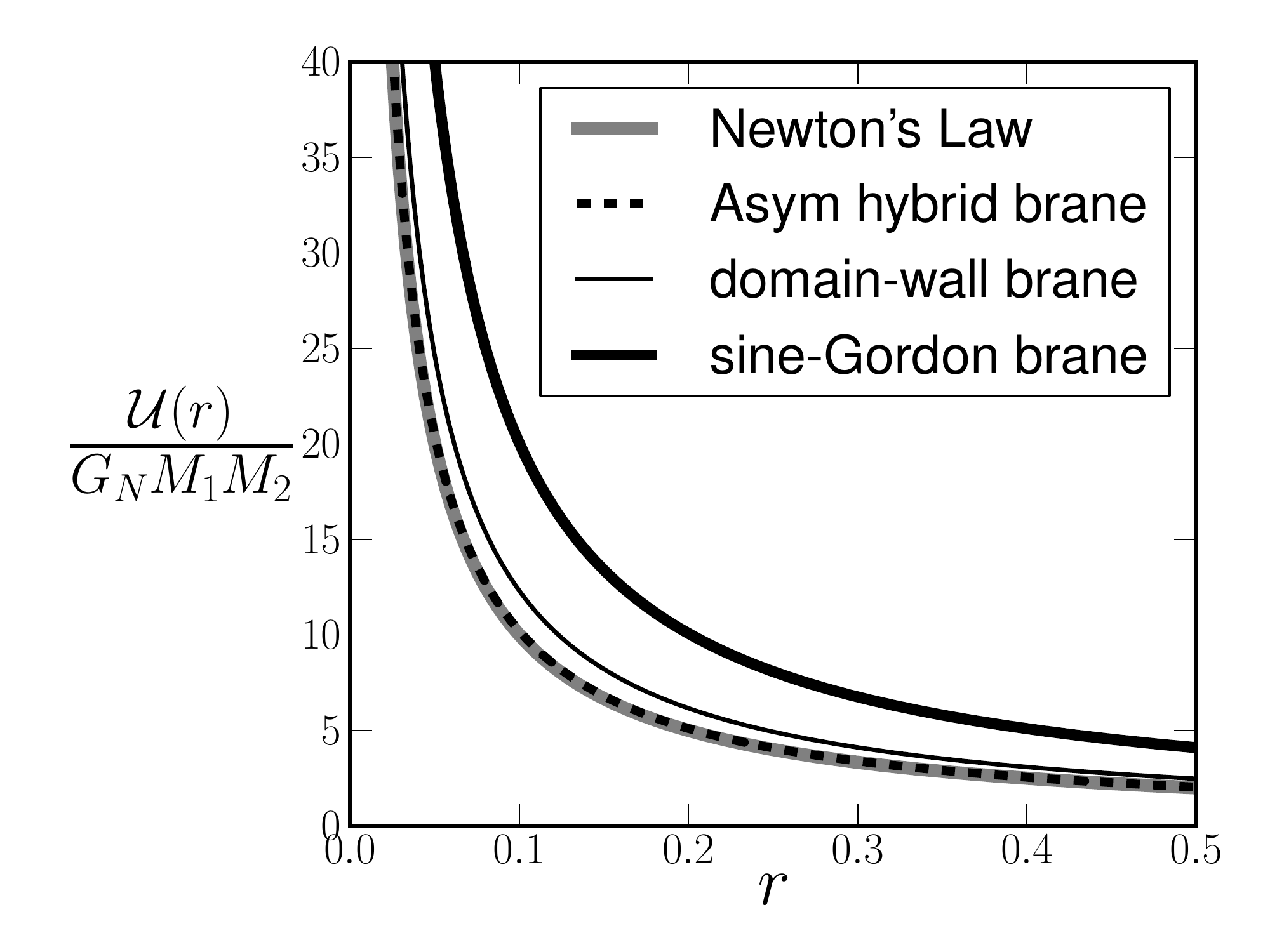}
 \caption{Newtonian potential with corrections due to Kaluza-Klein modes (thick gray line). The asymmetric hybrid brane (dashed black line) has a negligible correction. The sine-Gordon brane (thick black line) provides a higher correction in comparison to the ordinary domain wall brane (thin black line).}
 \label{Fig_Correction}
\end{figure}

We are now able to compute the corrections in the Newton's law. The procedure may be done for all the values of $p$ and $\lambda$, but we will concentrate ourselves in the cases shown in Fig. \ref{Fig_Spectra}, only. Moreover, from the Eq. \eqref{NewtonianConstant}, we are allowed to set $G_5 = {M_{*}}^{-3}$. Note that the Planck Mass will determine the spatial range of the interaction. In order to have a clearer understanding, we plot the corrected Newtonian potential in Fig. \ref{Fig_Correction} for the asymmetric hybrid brane, for the sine-Gordon brane and for the $\phi^4$ brane. The correction due to the compact brane is trivial. Moreover, the asymmetric hybrid brane has a negligible contribution. On the other hand, significant results are observed in the domain-wall and sine-Gordon branes. The sine-Gordon brane supports a larger mass spectrum providing a higher correction in comparison to the domain-wall brane. Therefore, we conclude that the sophisticate scenarios do not possess phenomenological implications, leaving the simplest thick braneworlds the trust of possible realistic models.

\section{Discussion}
\label{Sec-Discussion}

The phenomenology of the braneworld scenarios is parametrized by the first Kaluza-Klein (KK) graviton masses $m_n$. However, in general, the KK masses are determined by the characteristic radii of curvature of the geometry, so the spectrum may have complicated dependence on the internal geometry having no closed expression. Moreover, the parameters in the thick braneworld scenarios, related to the thickness of the defects, which in turn, are comparable to the bulk curvature, are arbitrary. Then, experimental measurements should be used to suit the parameters range. While this is not (\textit{yet}) tangible, theoretical manners to study the best set of parameters values are always welcome, such as the correction in the Newton's law of gravitation. For instance, recently, Correa \textit{et al} \cite{Config-Entropy} studied the bounds on the parameters of thick braneworld models in six dimensions using an information-entropic measure called configurational entropy. Using the approach developed in this paper, it can be inferred which braneworld should have a phenomenological result. 

The gravitational massless mode $\psi_0(z)$ is the first main quantity to study the building of a braneworld scenario. Its normalization reproduces the four-dimensional gravity. Furthermore, the normalizability of the zero-mode is intimately connected with the asymptotic behavior of the Schr\"odinger potential. If $U(z)\geq 0$ as $|z| \rightarrow \infty$, then $\psi_0(z)$ is always normalizable. On the contrary, if $U(z)<0$ as $|z| \rightarrow \infty$, then $\psi_0(z)$ is not normalizable and therefore, is of no interest, since it cannot describe four-dimensional gravity \cite{Csaki-UniversalAspects}. 

For the case when $U(z) >0$ as $|z| \rightarrow \infty$, the excited states are separated by a gap from the ground state \cite{Csaki-UniversalAspects}. Hence, corrections to Newton's law are exponentially suppressed as Eq. \eqref{Correction-Newton-Law}.  In the case where the potential goes to zero at infinity, there is a continuum of scattering states with eigenvalues $m^2 \geq 0$ and so the behavior of the light modes at $z = 0$ is crucial whether the corrections are small. However, in practice, the Schr\"{o}dinger-like equation for thick brane models can not be solved analytically, then a numerical procedure is needed. In such cases, the problem has to be imposed in a finite domain, which returns directly a discrete spectrum with a mass gap from the ground state.

The essential quantities to study the phenomenology of a braneworld scenario in a correction to the gravitational potential is the pair eigenvalue-eigenfunction of the Schr\"{o}dinger-like problem describing the massive states. The correction is exponentially suppressed in the mass and depends on the values of the wave functions in the origin ($z = 0$). We find that only odd-eigenfunctions contribute to the correction, since the value of $\psi(z = 0)$ is null for even-eigenfunctions.




\section{Conclusions and Perspectives}
\label{Sec_Conclusion}

In this work, we sought to draw attention to the analysis of the graviton massive states in codimension-$1$ warped braneworld scenarios. In addition to the localized massless mode, it is important to study the Kaluza-Klein (KK) tower, which conducts to corrections in the Newton's law of gravitation, the primary phenomenological implication of the braneworld hypothesis \cite{RS2}. We have studied the correction in the gravitational potential in the newest brane models proposed in the literature, the so-called asymmetric hybrid brane \cite{AsymmetricHybridBrane} and compact brane \cite{CompactBrane}. Since these models come from deformations of the $\phi^4$  and sine-Gordon topological defects, we also computed the correction in branes engendered by such defects. We have used suitable numerical methods to attain the KK spectrum and its corresponding eigenfunctions. As expected, the spectra are real and monotonically increasing. To test the numerical routines, we have applied the matrix method in the Randall-Sundrum (RS) model, and we have obtained good approximations for the first eigenvalues which are of physical interest. Moreover, we have reinforced the hybrid and the compact features of the models analyzing the scalar curvature and the analogue quantum potential.

We have shown that the compact brane only supports the massless state, then has no contributions to the correction in the Newton's law. This peculiar model corresponds to a braneworld scenario with high curvature. Moreover, the asymmetric hybrid brane has negligible contribution to the correction in the Newtonian potential. Therefore, the simplest braneworld scenarios carry the trust to be possible realistic models. The sine-Gordon brane provides a higher correction in the gravitational potential than the ordinary domain wall brane. We found that only odd-eigenfunctions contribute to the correction, since the values of the even wave functions are zero at the origin $z = 0$ (where the branes are supposed to be located). Thus, although all the spectrum accounts in the exponentially suppressed correction, the set of eigenfunctions selects the odd massive solutions only. Another important result from the numerical study of the massive eigenstates, is that the first normalized eigenfunction has the highest value at the origin, thus, effects of the Kaluza-Klein tower may be arising from the first excited state.

The procedure addressed in this paper is very useful to determine if a braneworld model should have phenomenological implications. This may also be used as a selection tool for braneworld models to match with future experimental measurements.

An extension of the current work consists in applying the techniques developed here in other braneworld models which present resonant modes, such as the Bloch branes \cite{Wilami-Resonance-Deformed, Wilami-Resonance-2field} and study the influence of the resonant Kaluza-Klein states in the correction of the Newtonian potential.

\section{Acknowledgments}

The authors thank the Coordena\c{c}\~{a}o de Aperfei\c{c}oamento de Pessoal de N\'{i}vel Superior (CAPES), the Conselho Nacional de Desenvolvimento Cient\'{i}fico e Tecnol\'{o}gico (CNPQ), and Funda\c{c}\~{a}o Cearense de apoio ao Desenvolvimento Cient\'{i}fico e Tecnol\'{o}gico (FUNCAP) for financial support.

\end{document}